%% file: msravi.tex
\documentclass{aa}  
\usepackage{graphicx}
\usepackage{txfonts}
\usepackage{epsf}
\input{bk_macros}

\def\dev{r^{1/4}~}
\def\ser{r^{1/n}~}
\def\sbe{{\mu_{\rm b}(\re)}}            
\def\msbe{\mmubre}                      
\def\mie{\mean{I_{\rm b}(<\re)}}        
\def\kmps{\unit{km \, s^{-1}}}
\def\magper{\unit{mag\, arcsec^{-2}}}
\def\ul{\underline}
\def\sqb #1{\left[#1\right]}
\def\sib #1{\left(#1\right)}
\def\cub #1{\left{#1\right}}
\def\oii {${\rm [OII]\lambda}$}
\def\refbf #1{{\bf{#1}}}
\newcommand{\plotn}[2]   
{\begin{figure}
\centering
\includegraphics[width=0.5\textwidth]{#1.eps}
\caption{#2}
\label{#1}%
\end{figure}}
\newcommand{\plotbig}[2]   
{\begin{figure*}
\centering
\includegraphics[width=0.9\textwidth]{#1.eps}
\caption{#2}
\label{#1}%
\end{figure*}}

\begin{document}
   \title{531 new spectroscopic redshifts from the CDFS and  a test 
on the cosmological relevance of the GOODS-South field}


\author{C. D. Ravikumar\inst{1,2},
M. Puech\inst{1},
H. Flores\inst{1},
D. Proust\inst{1},
F. Hammer\inst{1},
M. Lehnert\inst{3,1},
A. Rawat\inst{4,1},
P. Amram\inst{5},
C. Balkowski\inst{1},
D. Burgarella\inst{5},
P. Cassata\inst{6},
C. Cesarsky\inst{7},
A. Cimatti\inst{8},
F. Combes\inst{9},
E. Daddi\inst{10,12},
H. Dannerbauer\inst{11},
S. di Serego Alighieri\inst{8},
D. Elbaz\inst{12},
B. Guiderdoni\inst{13,2},
A. Kembhavi\inst{4},
Y. C. Liang\inst{14},
L. Pozzetti\inst{15},
D. Vergani\inst{1},
J. Vernet\inst{7},
H. Wozniak\inst{13}, 
X.Z. Zheng\inst{11}
}
\offprints{Ravikumar}

\institute{GEPI, Observatoire de Paris-Meudon, 92195 Meudon, France, 
\email{Chazhiyat.Ravikumar@obspm.fr}
\and 
Institut d'Astrophysique du CNRS, 98 bis Boulevard Arago, F-75014 Paris, France
\and 
Max-Planck-Institut f\"ur extraterrestrische Physik, Giessenbachstrasse, 
85748 Garching bei M\"unchen, Germany
\and 
Inter University Center for Astronomy \& Astrophysics, Post Bag 4,
Ganeshkhind, Pune 411007, India
\and 
Laboratoire d'Astrophysique de Marseille, Observatoire Astronomique de 
Marseille-Provence, 2 Place Le Verrier, 13248 Marseille, France
\and 
Dipartimento di Astronomia, Vicolo Osservatorio 2, I-35122, Padova, Italy
\and 
ESO, Karl-Schwarzschild Strase 2, D85748 Garching bei M\"unchen, Germany
\and 
INAF, Osservatorio Astrofisico di Arcetri, Largo Enrico Fermi 5, I-50125, Florence, Italy
\and 
LERMA, Observatoire de Paris, 61 Av. de l'Observatoire, 75014 Paris, France
\and 
National Optical Astronomy Observatory, 950 North Cherry Avenue, Tucson, AZ 85719
\and 
MPIA, K{\"o}nigstuhl 17, D-69117 Heidelberg, Germany
\and 
CEA Saclay/DSM/DAPNIA/Service d'Astrophysique, Orme des Merisiers, F-91191 
Gif-sur-Yvette Cedex, France
\and 
Centre de Recherche Astronomique de Lyon, 9 Avenue Charles André, 69561 Saint-Genis-Laval 
Cedex, France 
\and 
National Astronomical Observatories, Chinese Academy of Sciences, 20A Datun Road, 
Chaoyang District, Beijing 100012, China
\and 
INAF - Osservatorio Astronomico di Bologna, via Ranzani 1, 40127 Bologna, Italy}

\date{Received {date}; accepted <date>}

\abstract
{This paper prepares a series of papers analysing the
Intermediate MAss Galaxy Evolution Sequence (IMAGES) up to a redshift of
one. Intermediate mass galaxies ($M_J \leq  -20.3$) are selected from the
Chandra Deep Field South (CDFS) for which we identify a serious lack of
spectroscopically determined redshifts.}
{Our primary aim in this study is therefore to obtain a sample of
intermediate mass galaxies with known spectroscopic redshift to be used
for further analysis of their 3D-kinematics. We also intend to
test whether this important cosmological field may be significantly affected
by cosmic variance.}
{The spectroscopic observations were carried out using VIMOS on the ESO
VLT. The data reduction was done using a set of
semi-automatic IRAF procedures developed by our team.}
{We have spectroscopically identified 691 objects including 580
galaxies, 7 QSOs, and 104 stars.  The overall completeness achieved
is about 76\% for objects with $I_{AB} \leq 23.5$ in the CDFS after excluding instrumental failures. This
study provides 531 new redshifts  in the CDFS. It
confirms the presence of several large scale structures in the CDFS, which are found to be more prominent than in other redshift surveys .
To test the impact of these structures in the GOODS-South field, we construct a representative redshift catalogue of 640 galaxies with $I_{AB} \leq
23.5$ and z $\leq$ 1.
 We then compare the evolution of rest-frame U, B, V and K galaxy
luminosity densities to that derived from the Canada France Redshift Survey
(CFRS).  The GOODS South field shows a significant excess of luminosity densities in
the z=0.5-0.75 range, which increases with the wavelength, reaching up to 0.5 dex at 2.1 $\mu$m. Stellar mass and specific star formation evolutions
might be significantly affected by the presence of the peculiar large scale
structures at z= 0.668 and at z= 0.735, that contain a significant excess
of evolved, massive galaxies when compared to other fields.}
{This leads to a clear warning to results based on the CDFS/GOODS South
fields, especially those related to the evolution of red luminosity densities, i.e. stellar mass density and specific star formation rate. Photometric redshift techniques, when applied to that field, are producing quantities which are apparently less affected by cosmic variance (0.25 dex at 2.1 $\mu$m), however at  the cost of the difficulty in disentangling between evolutionary and cosmic variance effects.}


\authorrunning{Ravikumar \etal}
\titlerunning{Redshift survey in CDFS} 

 \maketitle
%

\section{Introduction}
Studies suggest a growing evidence of significant evolution in the
ensemble of galaxies out to $z=1$. Analysis of stellar population of
galaxies shows that the bulk of star formation occurs between $z=1$ and
$z=0.4$, mostly dominated by star formation in galaxies in the range
$3-30 \times 10^{10} \msol$ (Heavens \etal 2004; Hammer \etal 2005).
This increased star formation activity is reflected in the strong
evolution shown by galaxies in this redshift range from the analysis of
the $15\mu$m observations by ISOCAM (Elbaz \etal 2002).  These infrared
selected galaxies, with their high star formation rate, averaging $\sim
100 \msol/$yr, contribute significantly to the star formation history
(Flores \etal 1999; Le Floc'h \etal 2005 and the references therein). The
Luminous Infra-Red Galaxies (LIRGs) are up to 40 times more numerous at
$z\sim1$ than locally, and such a strong luminosity evolution has been
attributed to rapidly increased frequency of interactions ($(1+z)^{3-4}$
for $z\leq1$; see e.g. Hammer \etal 2005). However, major merging events
appear to drive the high star-formation rates of only about 30\% of
LIRGs (Flores \etal 1999; Zheng \etal 2004).
Interestingly, some of these star-forming galaxies appear to be large, massive
disk galaxies, thus emphasizing the important role of 
studying the dynamics of intermediate mass ($3-30 \times 10^{10} \msol$)
galaxies for understanding galaxy evolution.

Through the ESO large program ``Intermediate MAss Galaxies Evolution
Sequence (IMAGES)'' we intend to (1) establish the evolution of
mass-to-light ratio; (2) test the different physical processes leading to
the present day Hubble sequence; (3) detail the star formation history of
each individual galaxy; and (4) test the evolution of mass-metallicity
relation, angular momentum, size, and mass.  The analysis involves
combining GIRAFFE and FORS2\footnote{For a description of the instruments,
GIRAFFE \& FORS2, see http://www.eso.org/instruments/} observations,
which are being carried out,
of a sample of intermediate mass galaxies, selected mainly by their
rest-frame $J_{\rm AB}$ magnitude, and importantly, with the largest
multi-wavelength photometric data available.  The 3D spectroscopic
observations using GIRAFFE (\eg Flores \etal 2004) can provide detailed
kinematics (\eg Puech \etal 2006a) including accurate Tully-Fisher
relations (Tully \& Fisher 1977, see Flores \etal 2006). It can also
provide electron density maps of distant galaxies (Puech \etal 2006b),
thanks to the high spectral resolution ($R \sim 10000$) of the instrument
allowing the [OII]$\lambda\lambda$3726,3729 doublet to be easily resolved.
Besides this, we will derive the chemical abundances of galaxies using
observations with FORS2 (Appenzeller \etal 1998), taking advantage of its 
excellent red sensitivity. 

The sample of intermediate mass galaxies for our study IMAGES is selected
mainly by their rest frame $J_{\rm AB}$ magnitudes ($M_{\rm J}(AB) \leq
-20.3$) corresponding to stellar masses greater than $1.5 \times 10^{10}
\msol$.  Because the CDFS represents one of the most deeply surveyed at all wavelengths, we have chosen to carry out IMAGES in this field.  Unfortunately, in the CDFS, this galaxy population had 
insufficient numbers
of spectroscopic redshifts available. The VVDS (Vimos VLT Deep Survey,
Le F\`evre \etal 2004), FORS2 (Vanzella \etal 2005, 2006) and K20 (Mignoli \etal
2005) surveys were intentionally designed to focus on faint magnitudes
(typically I$_{AB} \leq 24$) in order to preferentially identify objects
at higher redshifts which results in essentially picking up fainter (and
hence low mass) objects in the redshift range ($0.4 \leq z\leq 0.9$) we
are interested in.  With the redshift survey discussed here, we intend
to increase the number of redshifts of relatively bright and emission
line galaxies.  To study the evolution
of galaxies in a particular field, one must pay careful attention to how
representative the field is, for example, in its redshift distribution.
The existence of one or several significant peaks in the redshift
distribution of the CDFS, indicating the presence of large
scale structures, might be problematic in interpreting the results when studying galaxy evolution, such as in GOODS or IMAGES programs.  

The structure of the paper is as follows:  In \S2 we describe the
observation and data reduction of low-resolution VIMOS spectra from
the CDFS and in \S3 we provide the general properties of the sample
for which we obtain redshifts. Further, we analyze the cosmological
relevance of GOODS in \S4 by constructing a representative spectroscopic
catalog paying attention to the selection biases involved.  Finally,
we provide a conclusion in \S5.  We adopt the $\Lambda -$CDM cosmological model ($H_0 = 70 \kmps \rm{Mpc}^{-1}$, $\Omega_M=0.3$
and $\Omega_\Lambda =0.7$).  Also, all magnitudes used in this paper
are in AB system, unless explicitly noted otherwise.

\section {Observations}
\labsecn{obs}
%
\plotn{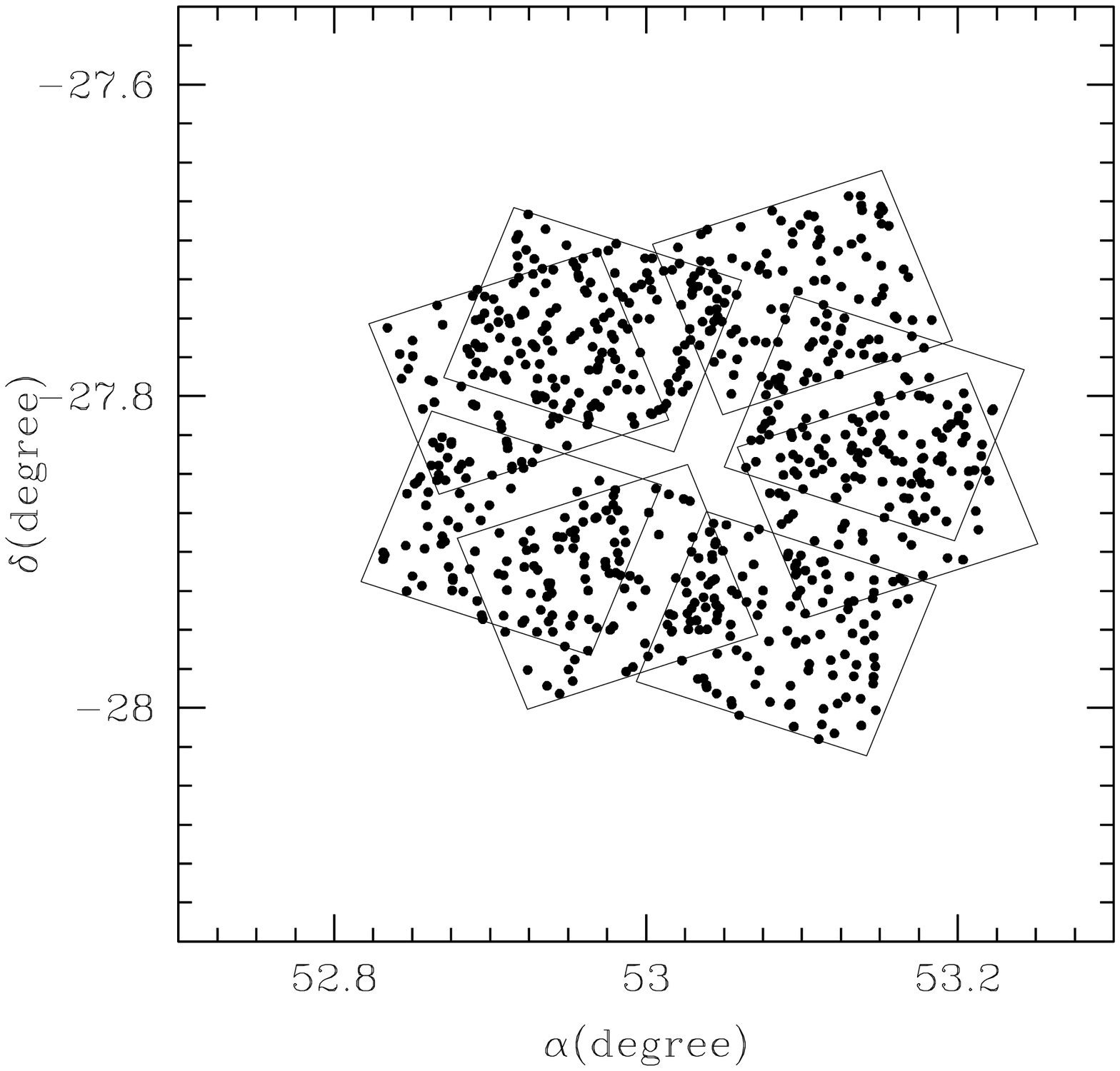}{The two pointings by VIMOS in the CDFS for observation with a similar strategy than VVDS. The
black dots are the positions of the objects for which we have determined
redshifts. A sample of 969 objects with $I_{AB} \leq 23.5$ were selected
for observation constituting $\sim 25\%$ of galaxies on the field.} %

We have carried out the spectroscopic observations with VIMOS on the VLT-UT3 Melipal during December 3-4, 2004.
The observing conditions were photometric and the seeing was $\sim$1
arcsec. The low resolution red grism, LRRED, was used with a slit width
of 1 arcsec.  The spectral resolution in this mode is $34\AA$ at $7500
\AA$ or $R\sim220$.  The red bandpass filter, which also serves for order
separation, limits the spectral range to $5500-9500\AA$. A complete VIMOS
pointing consists of observations with four quadrants of the instrument
each separated by about 2 arcmin from its neighbor. We have set two
pointings around the CDFS as shown in the Fig. \ref{area} with an angle
of $\sim90$ degrees between them.  The total area covered is $\sim 370
{\rm ~arcmin^2}$ on the sky. An integration time of 45 minutes was given
per pointing. Given the short exposure time, we have concentrated the
slits mainly on galaxies of the field with $I_{AB} \leq 23.5$ in order
to complement the existing deeper surveys.  The positions of the objects
for which we could estimate redshifts are also shown in Fig. \ref{area}.

The preparation of the slit masks was done using the
photometric catalog from ESO Imaging Survey (Arnouts
\etal 2001) and the I-band VIMOS pre-images.  The VMMPS
code\footnote{http://134.171.56.104/observing/p2pp/OSS/VMMPS/VMMPS-tool.html}
was run to optimize the number and positions of the slits for each
quadrant.  However, we have carried out a final re-arrangement of
the slits by hand in order to improve the number of objects observed.
This resulted in placing a total of 1142 slits in the two 
pointings, among them 969 were associated with objects with $I_{AB} \le 23.5$.


Data reduction and extraction of the optical spectra were performed
by three of us (CDR, HF and DP) using a set of semi-automatic
IRAF\footnote{IRAF is distributed by the National Optical Astronomy
Observatories, which are operated by the Association of Universities
for Research in Astronomy, Inc., under cooperative agreement with
the National Science Foundation.} procedures developed by our team.
The routines allowed us to reconstruct simultaneously the spectra of the
object and the surrounding sky. The spectra were flux-calibrated using
two standard stars. Images in the V, R and I bands were used to check
the slope of our spectra.

\subsection{Redshifts}
\plotbig{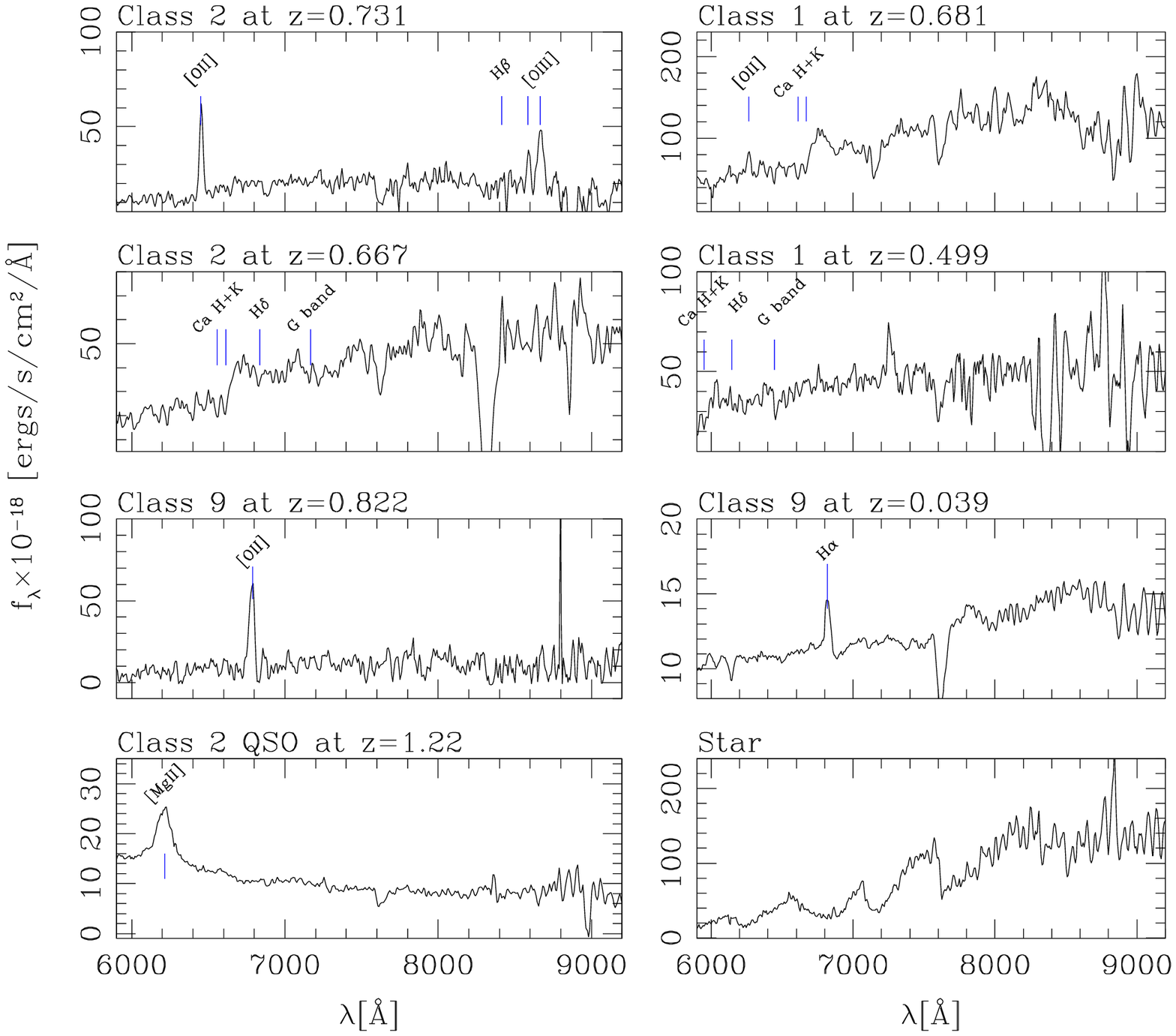}{Classification of spectra based on their quality. Sample spectra 
of various type and quality (Class 2: secure, class 1: insecure, class 9: single line). 
({\bf From top to bottom:}) emission line galaxies, absorption line galaxies, single 
emission line galaxies.  ({\bf Bottom:} AGN and star.}

The redshift determination of each spectrum from all the eight quadrants 
was done independently by three team members (CDR, HF \& DP), making sure that
each spectrum is analysed by at least two members.   
When the individual estimates of the redshifts among the three members
did not agree, measurements were examined by a fourth team member
(FH). When the fourth team member agrees with neither of the previous 
independent estimates, that object is discarded. Each redshift measurement
is associated with a `quality' flag denoting the reliability of the
extraction. Given the low signal-to-noise, mainly due to the short total
integration time, we have classified the
reliability into four classes; secure (class=2), insecure (1), single
emission line (9) and failed (0) observations. Secured spectroscopic
identifications contain more than two strong features, while sources
classified as insecure contain either a strong feature with not very
reliable supporting features, or multiple features not strong enough
to confirm the redshift, giving a confidence of about only 50\% to the
estimated redshift. Class (9) sources correspond to spectra with a single strong emission line without
any other features.  Redshifts are assigned tentatively for such cases.
Typically this is the case where we are unable to identify between \oii
3727 and ${\rm H{\alpha}}$ emission lines. The failed class contains
spectra for which we could not  obtain a redshift, mainly due to
low signal-to-noise  or to reasons that have an instrumental origin,
like the presence of bad pixels or object being near the boundary of
the slit. Figure ~\ref{figspecravi} shows the sample spectra for the different
classes used. In addition to this we have provided each redshift with a
type classification to show the observed spectrum belongs to an emission
line galaxy (type=1), an absorption line galaxy (2), a quasar (3), or a
star (4).

\begin{table}
\begin{center}
\caption{Number and type of objects in the classes, secure (2), insecure 
(1) and single emission line (9), for which we have determined redshifts. 
The \% is the percentage of objects for which redshifts are obtained out 
of 1142 slits placed.  For about 12\% of objects we were not able to 
determine the redshift because of instrument-related problems. 
See text for more details.}
\label{tab0}
\begin{tabular}{cccccc}\hline\hline
Class &Galaxy & Star & QSO & Total & \% \\
\hline
2  & 308 & 88 & 6 & 402 & 35.2 \\
1  & 206 & 16 & 1 & 223 & 19.5 \\
9  & ~66 & -- & - & ~66 &  5.8 \\
\hline
\multicolumn{4}{c}{Total}& 691 & 60.5 \\
\hline
\hline
\end{tabular}
\end{center}
\end{table}
 
\subsection{The redshift catalog}
The catalog lists spectroscopic data on 691 objects from the CDFS, containing 
580 galaxies, 104 stars and 7 QSOs (see Tables 1 and 2). The catalog 
can be downloaded from the CDS website.  In the catalog we provide for each
object:
\begin{itemize}
\item The ESO imaging survey (EIS) identification number and equatorial coordinates
(equinox 2000);
\item The class and type codes denoting the quality and nature of the spectra,
as described in the previous section;
\item The redshift measured by the IMAGES team;
\item The internal identifier;
\item The isophotal $I_{AB}$ band magnitude from Arnouts et al. (2001), and the derived absolute magnitudes (AB system) following Hammer et al. (2005) on the basis of available EIS photometry;  $M_{U}$ values have been calculated on the basis of observed V and V-I to allow comparison with CFRS values (see section 4.3), and then might be affected by systematics at low redshift.
\end{itemize}

\section{Description of the IMAGES catalog}

Out of the 1142 slits placed, we could estimate redshifts for 691
objects, giving a success rate of $\sim 61$\%.  Taking into
account the 969 objects with $I_{AB} \leq 23.5$, we successfully estimate the redshift of 635 of them ( success rate of 66\%).  The failure to estimate a redshift is associated mainly to
low signal to noise ($\sim 26$\%), objects falling near the edge of the
slit ($\sim11$\%), and presence of column of bad pixels ($\sim2$\%).

We have observed 12 galaxies twice.  By taking the difference of the
two individual independent redshift measurements, we find a systematic
uncertainty of $0.006 \pm 0.014$ in our redshift estimates.  With the
limited statistics available, we estimate a very conservative upper-limit
to the dispersion in our redshifts of 0.010.

\plotn{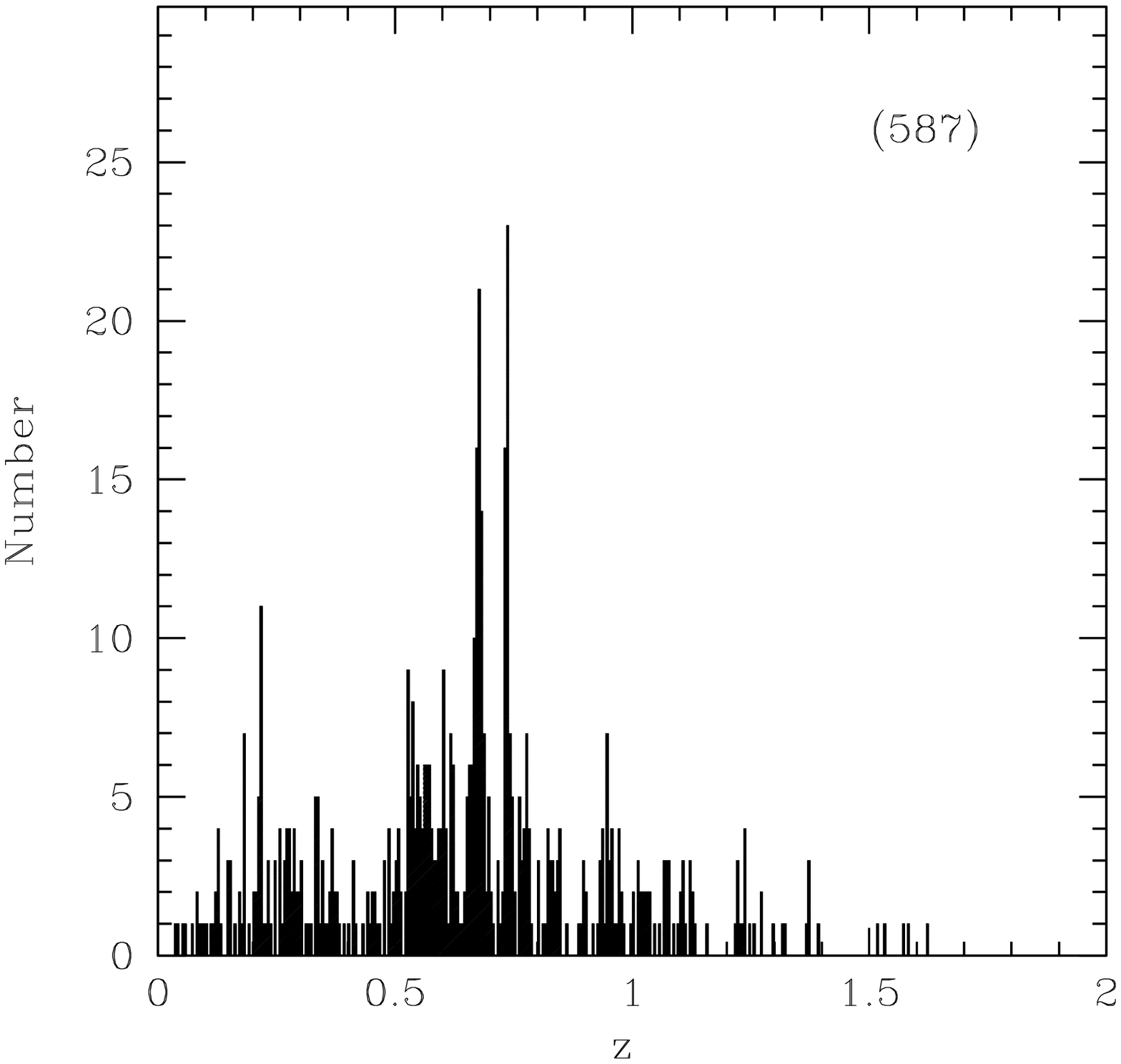}{Histogram of redshifts for galaxies and QSOs in our sample, 
comprising of all classes 2,1, and 9. The size of the redshift bin 
is 0.005. The distribution shows prominent peaks at 0.210, 0.530, 0.670, 
and 0.735.}

In Fig. ~\ref{allz} we show the distribution of the redshifts for all
galaxies and QSOs in our sample, consisting of all classes 2, 1, 
and 9.  The mean and median of the distribution
are 0.643 and 0.665 respectively.  The strongest peaks in redshift are
at 0.670 (82 objects, with $\Delta z=\pm0.020$), at 0.735 (58) and at
0.210 (23). The first two peaks are in agreement with the existence of
`wall-like' structure in the CDFS (Le F\`evre \etal 2004; Gilli \etal
2003, Vanzella \etal 2005).

\subsection{Completeness}
\setcounter{table}{2}
\begin{table*}
\begin{minipage}[h]{\textwidth}
\renewcommand{\footnoterule}{}  
\begin{center}
\caption{Comparison of the number (N) of galaxies (and AGN) with redshifts available in CDFS as a 
function of $I_{AB}$ limiting magnitudes. Our study represents from 20 to 35\% of galaxy redshifts available in the CDFS. The last column shows the number of galaxies observed in common from one survey to the others, based on the cross correlation of targets in the CDFS. 136 galaxies and 24 stars in IMAGES are found in common with other surveys, leaving us with 587-136=451 unique identifications of galaxy redshifts and 691-160=531 unique objects in IMAGES.}
\label{tab1}

\begin{tabular}{|lcccccc|}\hline\hline
Sample &Time\footnote{Approximate average exposure time per slit in minutes. The higher 
integration times used by other studies were justified by their scientific goals.}  
& $I_{AB} \leq 22.5$ & $I_{AB} \leq 23.5$ & $I_{AB} \leq 24$ & All & Common \\
\hline
K20     & -   & ~169 & ~236 & ~256 & ~459 & 133 \\
VVDS    & 265  & ~552 & 1089 & 1460 & 1460 & 223 \\ 
FORS2   & 281  & ~127 & ~219 & ~308 & ~711 & 144 \\
Szokoly & 114  & ~~77 & ~~89 & ~~90 & ~121 & ~55 \\ 
IMAGES  & ~45  & ~376 & ~532 & ~569 & ~587 & 136 \\
\hline
\multicolumn{7}{|c|}{Unique Identifications}\\
\hline
ALL     & - & 1068 & 1846  & 2328 & 2781 & - \\ 
IMAGES  & - & ~280 & ~401  & ~435 & ~451 & - \\
\%\footnote{The percentage of galaxies with new redshift identifications from the IMAGES survey. 
Calculated as IMAGES/(ALL - IMAGES).}      
& - & 35.5 & 27.8  & 23.0 & 19.4 & - \\
\hline
\hline
\end{tabular}
\end{center}
\end{minipage}
\end{table*}
\plotn{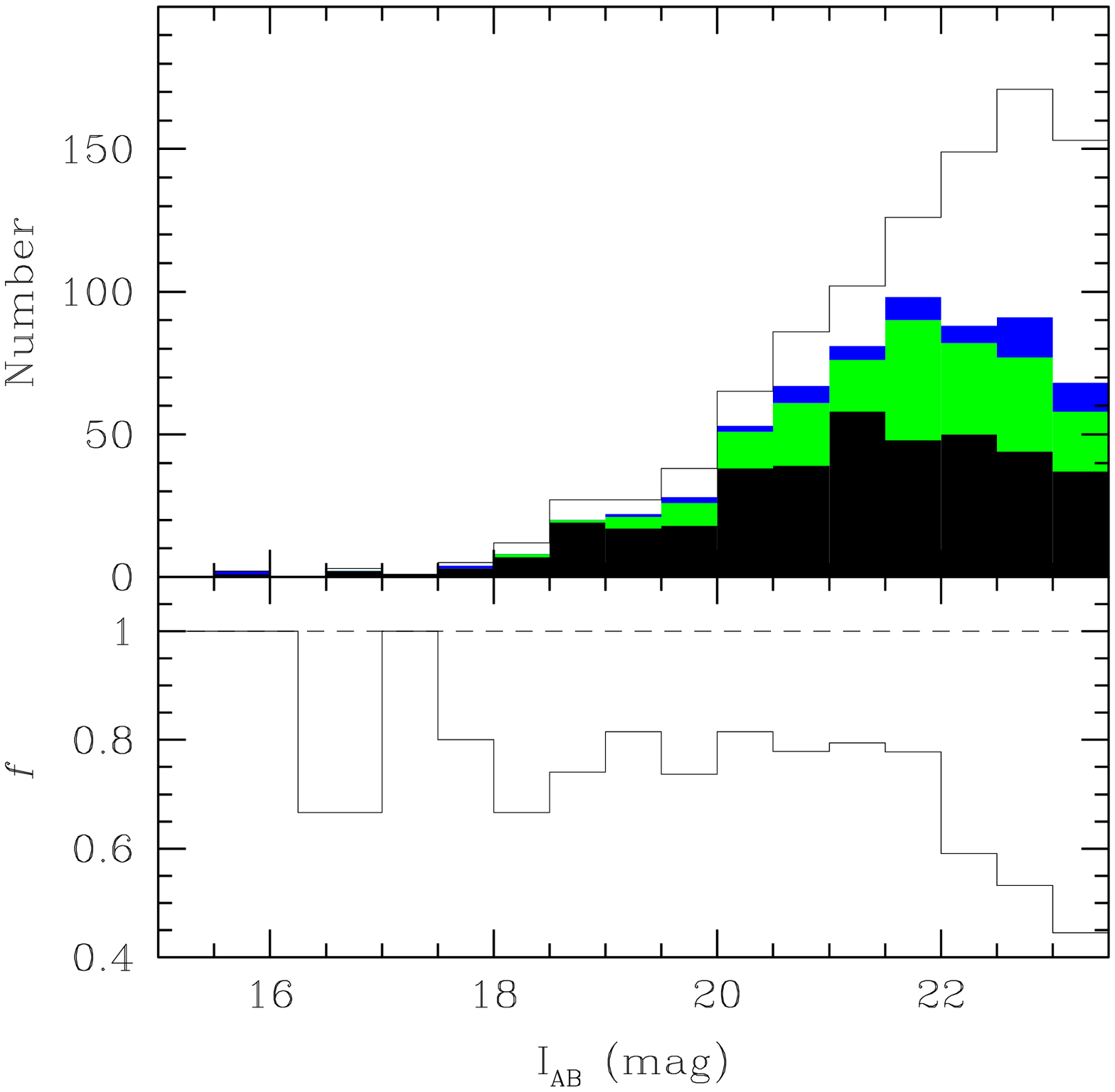}{Redshift measurement completeness for  $I_{AB} \leq 23.5$
sample in IMAGES. In the upper panel, we show the magnitude distribution
of objects with a redshift class 2 (black), 1 (green), and 9 (blue),
along with that for the observed targets (open). The lower panel shows
the histogram of the fraction ($f$) of objects with redshifts compared
to the total observed from IMAGES. We measure redshifts for 66\% of the
target objects, while the completeness achieved is 76\% when we exclude
failures that have an explicit instrumental origin.}

The spectroscopic completeness, i.e. the ratio of the actual number
of spectroscopic identifications obtained to that of total number
of objects observed spectroscopically as a function of magnitude,
helps in understanding the magnitude dependent bias of the sample.
As mentioned previously, for objects with $I_{AB} \leq 23.5$, the overall
spectroscopic completeness is 66\%.  If we exclude the failures that have 
an instrumental origin, then the  spectroscopic completeness
for $I_{AB} \leq 23.5$ objects is  76\%. In Fig. ~\ref{comp},
we show the spectroscopic completeness achieved in measuring redshifts
as a function of the apparent magnitude ($I_{AB}$). In the upper panel the
histogram of observed targets with $I_{AB} \leq 23.5 $ are shown in
white, while the black, green and blue histograms show the magnitude
distribution of objects with redshift class 2, 1 and 9, respectively.
Further, in Table 3, we show a comparison of the number of
galaxy redshifts determined by IMAGES with the K20, VVDS, FORS2, and Szokoly \etal (2004). 
 We have taken into
account only reliable redshifts, with a confidence of at least 50\%.
This means that redshift estimations with  flags 1, 2 \& 9 from IMAGES,
flags 1-4 \& 9 from VVDS, flags 0 \& 1 from K20, all from FORS2\footnote{Note that the 
flags used by Vanzella \etal are not exactly similar to that used by other 
surveys., and flags 0.5, 1, 2 \& 3 from Szokoly \etal 
were selected in Table 3. We note that our study (IMAGES) represents approximately 28\% of the 
reliable redshifts for galaxies with $I_{AB} \leq 23.5 $ in the CDFS. 
Excluding redshifts of objects already identified by other surveys (repeated observations, see Table 3), the IMAGES catalogue produces 531 new 
redshift identifications, consisting of 307, 170 and 54 objects with class codes 2, 
1, and 9.  The numbers of galaxies, stars and quasars present in this catalog are 
447, 80 and 4, respectively.}.
The high efficiency coupled with the 
relatively high completeness attained over a short exposure time highlights 
the advantage of low resolution spectroscopy by VIMOS in estimating redshifts 
up to $z=1.5$.

\plotn{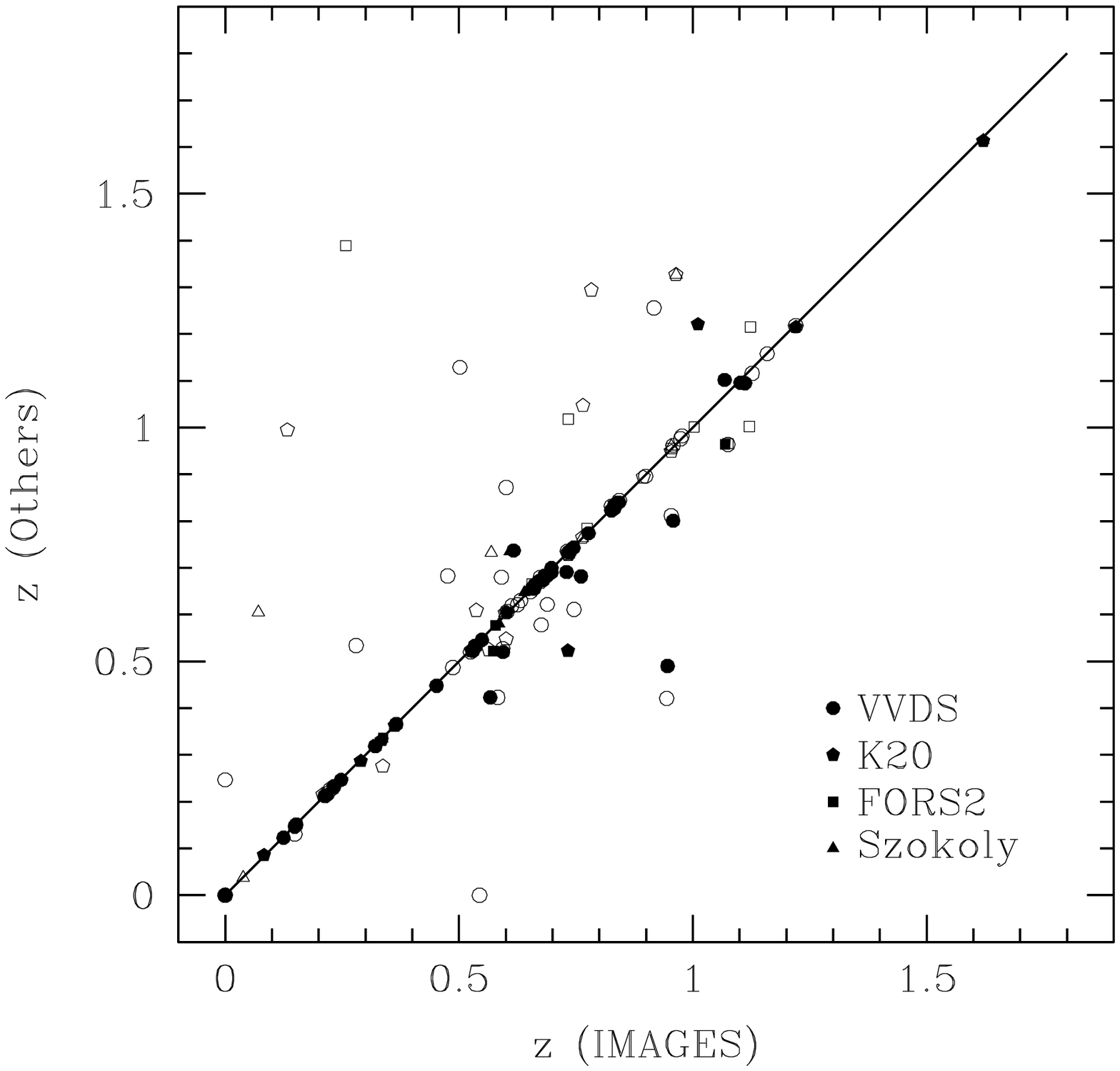}{The comparison of our redshift estimation with VVDS (open 
and shaded circles), K20 (open and shaded pentagons), FORS2 (open and shaded 
squares), and Szokoly (open and shaded triangles).  The open symbols identify objects for which at least one of the survey considers the redshift as insecure (for example, with flags 1 and 9 in both IMAGES and VVDS), otherwise a full symbol is adopted.
}

\subsection{Comparison with other redshift estimations in the CDFS}
We have compared our redshift estimations with those publicly 
available in the CDFS, i.e., the VVDS (Le F\`evre et al, 2004), K20 (Cimatti et al, 2002), FORS2 (Vanzella \etal 2005, 
2006) and the X-ray selected sample of Szokoly \etal (2004).  We have 
observed 98, 42, 22, and 18 objects in common to the VVDS, K20, FORS2 
and Szokoly \etal samples, respectively. Because former surveys also include target duplications, we find that 160 IMAGES objects were already observed in former surveys. For 76\% of the common elements with 
VVDS, we have $\frac{\Delta z}{(1+z)} \leq 0.01$. The distribution of 
redshift difference for 56 objects where both VVDS and IMAGES mark a 
secured estimation, shows $\sigma_{\Delta z} = 0.069 $. We obtain 
similar percentages and dispersions for other samples also. In 
Fig.~\ref{zcomp} we show the comparison of our redshift estimation with
VVDS (circle), K20 (pentagon), FORS2 (square)  and Szokoly \etal (triangle) 
samples. The shaded and open symbols are used to represent secure and 
insecure estimations.
Clearly, all the significant differences are arising from low quality of 
the spectra in at least one of the surveys.  Two of us (CDR \& HF) compared 
the spectra of objects
for which we obtained significantly different redshifts from VVDS, using
the CENCOS database\footnote{http://cencosw.oamp.fr/FR/index.fr.html}.
After comparison, we have kept our redshift for about 60\% of them.
When spectral features did not match with the VVDS ones (all with
very poor quality), we preferred our estimates because it appeared
likely that the features in the VVDS spectra are artificial and arise from
a failure by their automated data reduction in handling accurate sky
subtraction and/or wavelength calibration.  This was always possible
anyway, since the VVDS team themselves assigned the sources that we
find to have discrepant redshifts as of very poor quality. The redshift
estimation for the object EISJ033235.64-274633.0 is noteworthy. We
estimate a redshift of 0.564 while the VVDS team reports 3.6664 (GOODS
identification J033235.62-274632.8) with their quality flag 2. We find the
\oii 3727 emission at 5837.9 \AA, though with a class=1 as no more strong
features are observed. The $B$ band detection of the object ($B=24.11$)
also suggests an absence of the Lyman break feature corresponding to
$z = 3.6664$, thus supporting our estimation of the lower redshift.
This just highlights that caution should be observed when using redshift
estimates of marginal or poor quality.

\subsection{Stars and galaxies in the color-color diagram}
\plotn{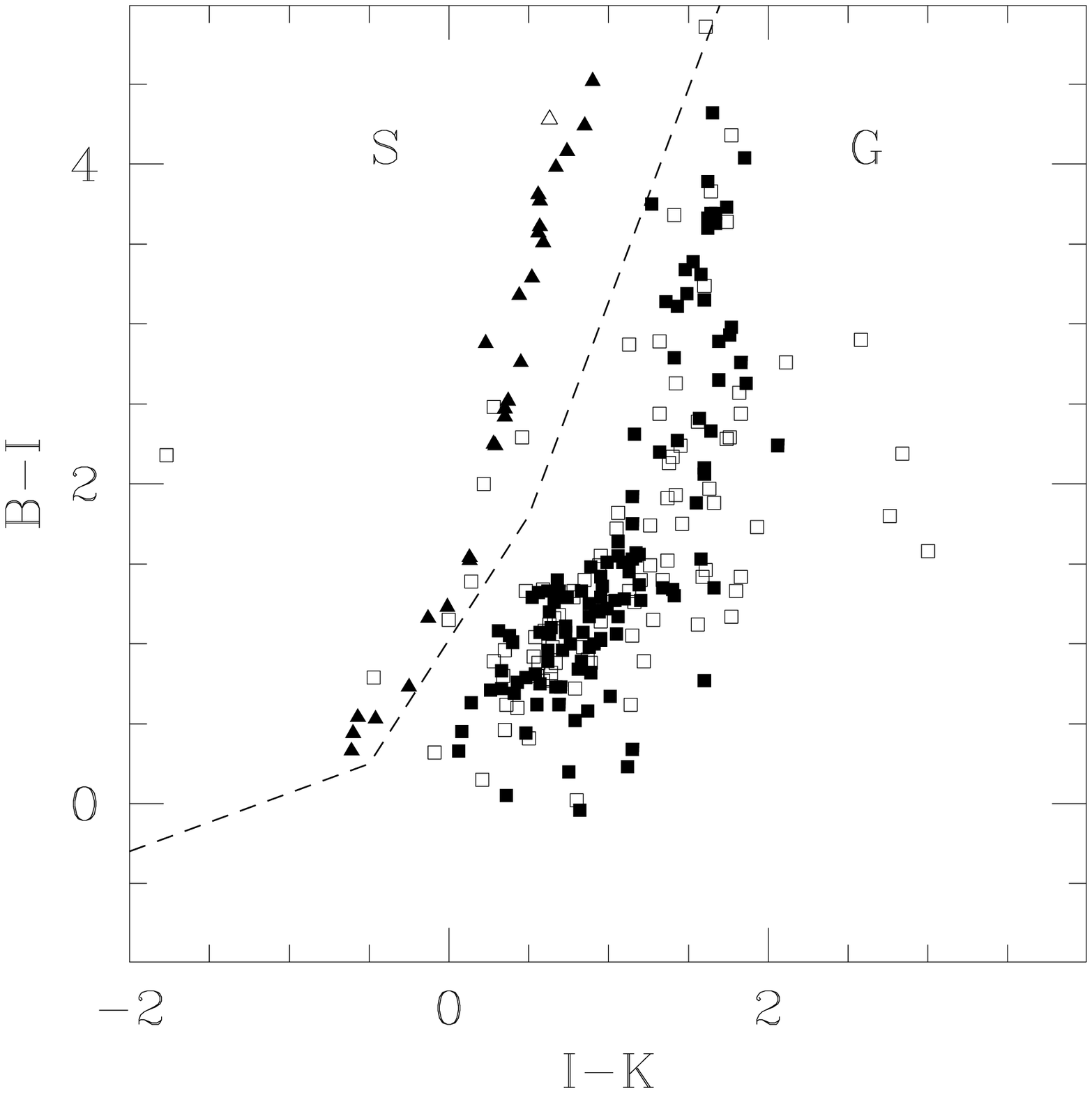}{The I-K vs. B-I diagram for stars (triangles) and galaxies
(squares) with spectroscopic redshift measurements from IMAGES. Filled
and open symbols represent secure and insecure redshift estimates
respectively. The dashed line delineates roughly the regions occupied
by stars (S) and galaxies (G).}
The color-color diagram, provides an easy tool in identifying possible
mis-classifications of stars and galaxies. However, there may be an overlap
between the colours of stars and galaxies, especially for the bluest objects (e.g., Crampton \etal
1995). In Fig. \ref{BIvsIK},  the variation of B-I as a function of I-K is
shown. Filled triangles and squares show stars and galaxies with secure
spectroscopic identifications, while the open symbols represent the same
classes for objects with less secure identifications.  Using the diagram,
we selected objects for visual examination, wherever possible, through
the available HST-ACS  and the GEMS images. We noticed 26 (4\% of total
objects with spectroscopic redshift) obvious cases of galaxies identified
as stars (insecure, with class 1) in the preliminary catalog. Due to this
uncertainty induced by inspecting the images, the quality flag of these
objects were degraded to unidentified (class 0) in the final catalog.

\subsection{Magnitude distribution}

The variation of the apparent I magnitude as a function of redshift for the
galaxies in our IMAGES catalog is shown in Fig. ~\ref{ziim}. Filled
squares represent secure redshift identification while the insecure
redshift estimates are shown by open squares. We have estimated the
absolute magnitudes following the method outlined in Hammer \etal (2001
\& 2005).  The distribution of absolute magnitude ($M_B$) versus redshift
is shown in Fig. \ref{MBdez}, indicating the absolute magnitude depth attained by
our observations. For clarity, the brighter ($I_{AB} \leq 23.5$) and
fainter ($I_{AB} > 23.5$) objects are represented by filled and open
circles respectively.

\plotn{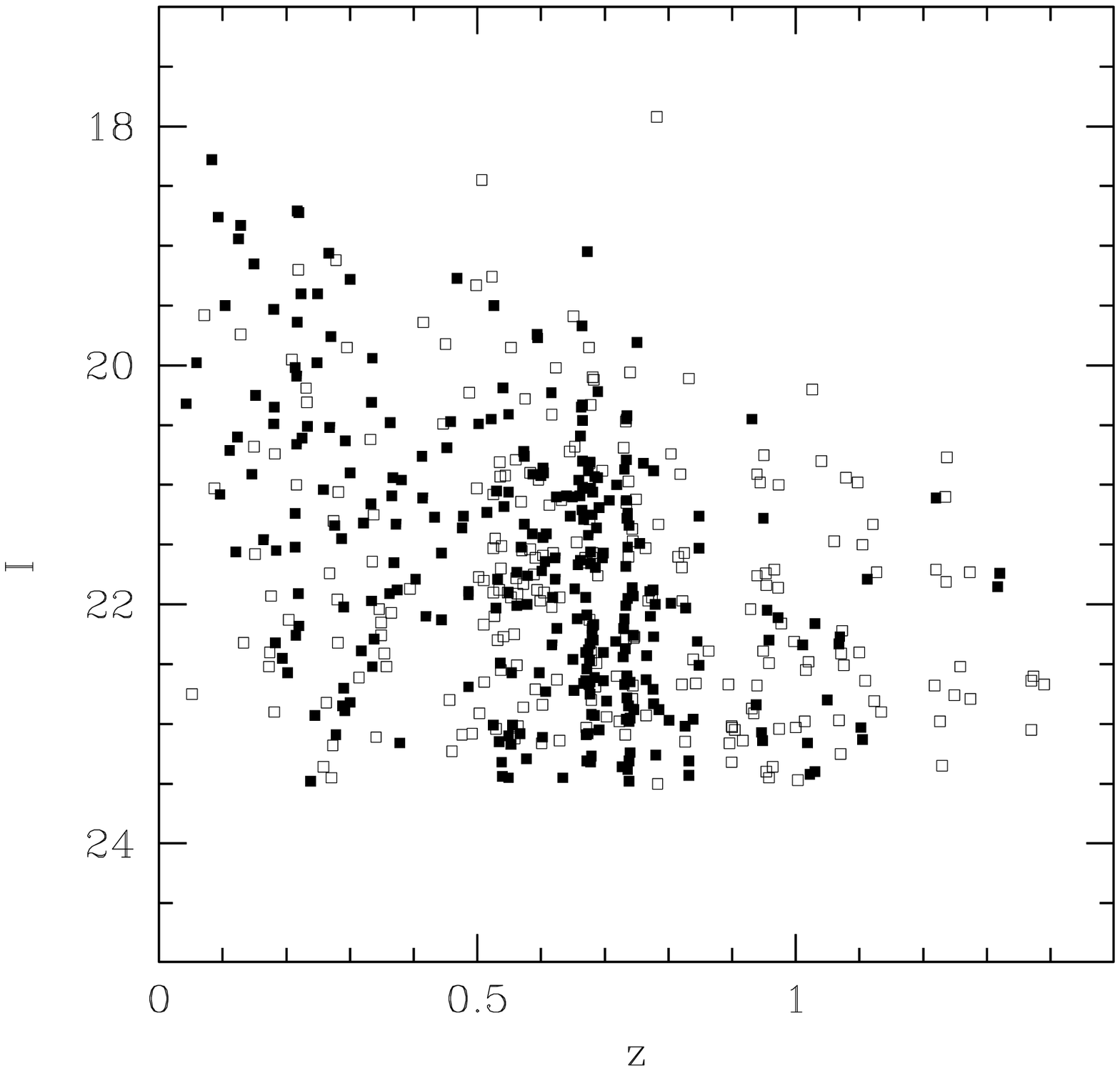}{The variation of the apparent I magnitude as a function of
redshift, for galaxies for which redshifts are measured by the IMAGES
survey. Filled and open squares show secure and insecure redshift
estimates respectively.}

\plotn{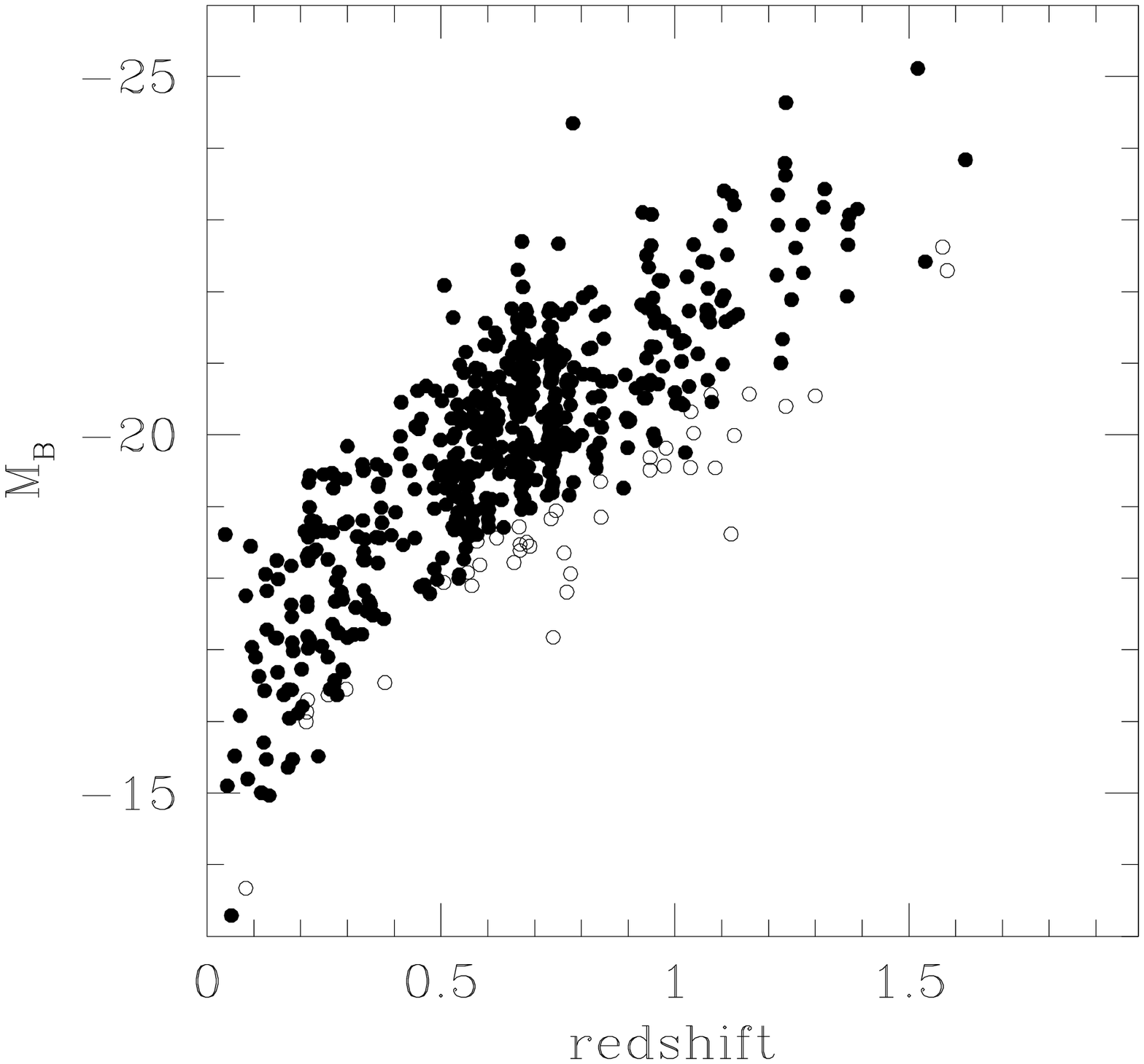}{Redshift vs absolute $\rm M_B$ magnitude for objects in
our sample showing the depth attained by our observation.  Objects
brighter and fainter than $I_{AB} = 23.5$ are shown as filled and open
circles respectively.}

\plotn{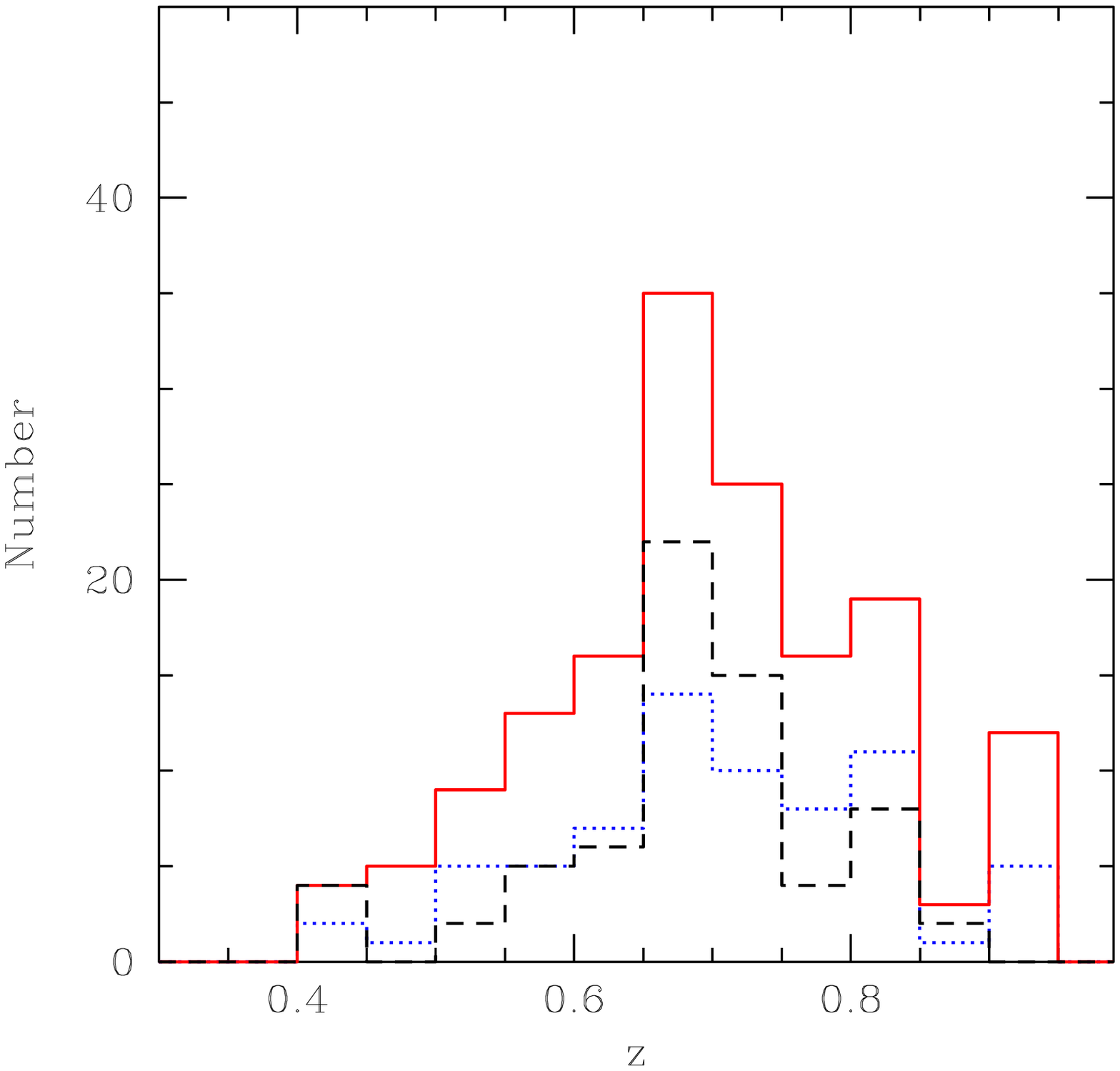}{Sample of galaxies that are candidates for GIRAFFE follow-up observations.
  The blue dotted histogram shows \oii 3727 emission
galaxies with $M_J \leq -20.3$, identified from IMAGES survey while the
black dashed histogram shows that from VVDS. The histogram with continuous
red line shows the distribution of emission line galaxies with $M_B \leq
-20.0$ but with no $M_J$ data available.  See text for more details.}

\subsection{The sample of emission line galaxies in the IMAGES survey}

One of the main criteria for selecting our sample of intermediate
mass galaxies, is the absolute magnitude selection $M_J \leq -20.3$.
Such a limit corresponds to a stellar mass of $1.5 \times 10^{10} \msol$
when converting the J band luminosity using the prescription discussed
in Bell et al (2003; see also Hammer et al, 2005). However, the near
infrared observations in $J$, $H$, and $K$, essential for estimating
$M_J$ using the method of Hammer \etal (2005), are currently available
only for the GOODS area, which forms a subset of the larger CDFS. But
we find for galaxies in our sample with both B and J magnitudes, about
95\% of $M_B \leq -20.0$ galaxies have $M_J \leq -20.3$. We obtained very
similar percentage with a much larger sample of galaxies with photometric
redshift (from COMBO-17, see Wolf et al, 2004).  The latter test was conducted in order to verify that the
selection biases involved in our sample (also see \S~\ref{rep}) do not
affect the fraction of $M_J \leq -20.3$ galaxies that could have $M_B >
-20$. Hence for emission line galaxies falling outside the area covered
by the GOODS, we select our sample using $M_B \leq -20.0$ criterion. In
Fig. ~\ref{mjhist}, we show the histograms of emission line galaxies
available in CDFS. The IMAGES survey identifies 69 galaxies with $M_J
\leq -20.3$ and \oii 3727 emission lines observable through GIRAFFE (blue
dotted line), while 68 are available from the VVDS survey (black dashed
line). We also identify 157 emission line galaxies with $M_B \leq -20.0$
for further observation, thus providing large enough sample of target
galaxies for detailed analysis, as envisaged in the IMAGES program.

In what follows we provide a representative sample of galaxies with
spectroscopic redshift in the GOODS field to address the issues related
to the possible cosmological variance within this field.

\section{The Representativeness of GOODS-South for Cosmological Studies}

The GOODS project envisages the use of the best and probably the deepest
multi-wavelength data in order to understand various cosmological
problems, like the formation and evolution of galaxies and active
galactic nuclei, and the distribution of dark matter and the large scale
structure at high redshifts. However, after this region was chosen,
it has been discovered that it had a ``wall-like structure''
(Le F\`evre \etal 2004; Gilli \etal 2003, Vanzella \etal 2005).
Finding such a correlation in the galaxy distribution necessitates a
more detailed analysis of issues related to the cosmological variance
that might impact our understanding of galaxy evolution when using data
from the CDFS region.  To conduct this investigation, we constructed
a sample of galaxies with reliable spectroscopic redshifts in the
GOODS area by combining the IMAGES data with other relevant publicly
available spectroscopic redshift surveys.  The sample is then used to
compare the global properties of galaxies observed in the CFRS. The CFRS
(see Crampton \etal 1995)
is a collection of data and analysis on galaxies with $I_{AB} < 22.5$,
selected in an unbiased way from 5 distinctly uncorrelated areas of 100 $arcmin^{2}$
in the sky to mitigate against the effects of cosmic variance (see Crampton \etal Figure 8 showing the redshift distribution of the 5 individual fields). This,
though significantly less deep than the observations in the GOODS,
the CFRS provides an ideal sample to test the possible impact of cosmic
variance in the GOODS region.

\subsection{The GOODS - South field spectroscopic catalogue: 640 galaxies
with $I_{AB} \le 23.5$ and $z \le 1$}\label{rep}

Though the IMAGES survey was highly efficient in estimating redshifts,
the short integration time of the survey introduces a real bias
towards brighter galaxies. This is evident in Fig. \ref{comp} as the
fraction of redshift identified objects becomes low at fainter magnitudes
($I > 22$). Moreover, the high efficiency of the IMAGES survey is
mainly reflected in its ability to obtain redshifts for galaxies with
strong emission line spectra.  The short integration time results in
low signal-to-noise for faint objects which in turn causes serious
difficulties in estimating redshifts of galaxies showing absorption
line only or weak emission lines.  Thus we tend to be particularly
biased against obtaining redshifts the fainter early-type galaxies.
On the other hand, we find that the selection criteria employed by the FORS2 team
 lead to spectroscopically identify those objects that are largely
missed by IMAGES.  The FORS2 spectroscopic team generally selected redder and
fainter galaxies significantly helping to overcome the biases in our
spectroscopic survey.  The VVDS catalog forms a more complete sample
(84\% completeness for objects with $I_{AB} \le 24$).

Here we limit our analysis to GOODS-South galaxies with $I_{AB} \le 23.5$ and $z \le 1$, with the aim of testing the cosmological relevance of the GOODS-South field within that redshift range. Fig. \ref{histo}  compares
properties of galaxies in VVDS (337 galaxies) to those of the combined sample of IMAGES
and FORS2 (303 galaxies). It reveals a remarkable consistency between the two samples, which photometric, color and redshift distributions are almost indistinguishable one from the other on the basis of statistical tests. In the following we combine the two catalogues, assumed to provide a representative sample of $z \le 1$ galaxies up to  $I_{AB} = 23.5$. In the following, the GOODS - South field spectroscopic catalogue corresponds to the combination of the VVDS and of IMAGES and FORS2 catalogues in the GOODS field. 

\plotn{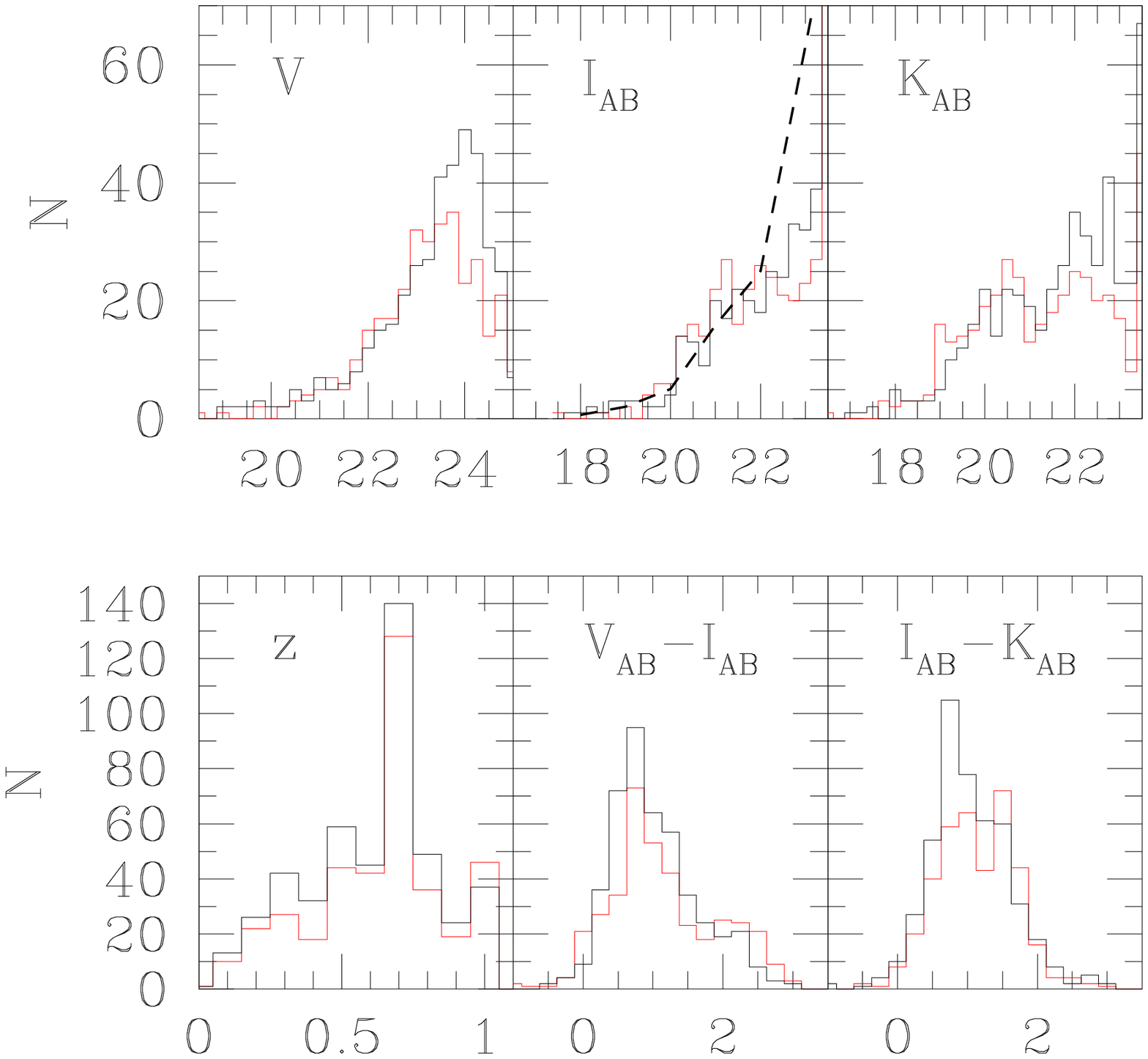}{Comparison between the distribution of VVDS (black histogram) and IMAGES+FORS2 (red histogram) galaxies in the GOODS field. From left to right, top to bottom, the two distributions are compared in V, $I_{AB}$, $K_{AB}$, z, $(V-I)_{AB}$ and $(I-K)_{AB}$, respectively. Kolmogorov-Smirnov tests show that the associated probabilities (for the two samples to be derived from the same population) are 96, 99.999, 82, 78, 96 and 99.999\%, respectively. Notice that a Pearson test provides similar numbers (probabilities always larger than 91\%). The dashed line in the top-middle panel represents the counts in the I band (see Arnouts et al, 2001), showing that both catalogues are missing galaxies at the faint end.}

%
  


\subsection{Comparison of the GOODS-S redshift distribution to that of GOODS-N, VVDS02h and CFRS}

We have retrieved from the literature existing redshift surveys of similar depths than our GOODS-S sample.  Fig. \ref{histo_z} presents the 4 redshift histograms, revealing in three cases (GOODS-S, GOODS-N and VVDS02h) the existence of large scale structures. As noticed above, the CFRS is a blend of 5 different fields of view, so the combination has smeared the large scale structures. For each field we identified the two most prominent structures (in yellow), and identify the redshift area (in blue) surrounding the structure, which contains the same number of galaxies than the two structures. This is similar to an "equivalent (redshift) width" revealing the impact of the structure in a given redshift survey. The two prominent structures in GOODS-S, GOODS-N and VVDS02hr provide equivalent (redshift) widths of $\Delta$z= 0.317, 0.148 and 0.09, respectively. It is unlikely that these differences can be related to the (small) differences in limiting magnitudes. Knowing that many studies of the evolution of integrated quantities such as SFR or stellar mass densities are considering redshift bins of $\Delta$z from 0.25 to 0.5, this clearly requires more investigations to understand if GOODS-S field can be biased by the presence of such prominent large scale structures. In the following, we will test their impact in deriving luminosity density, by comparing with the CFRS, which is by construction, not affected by structures.

\plotn{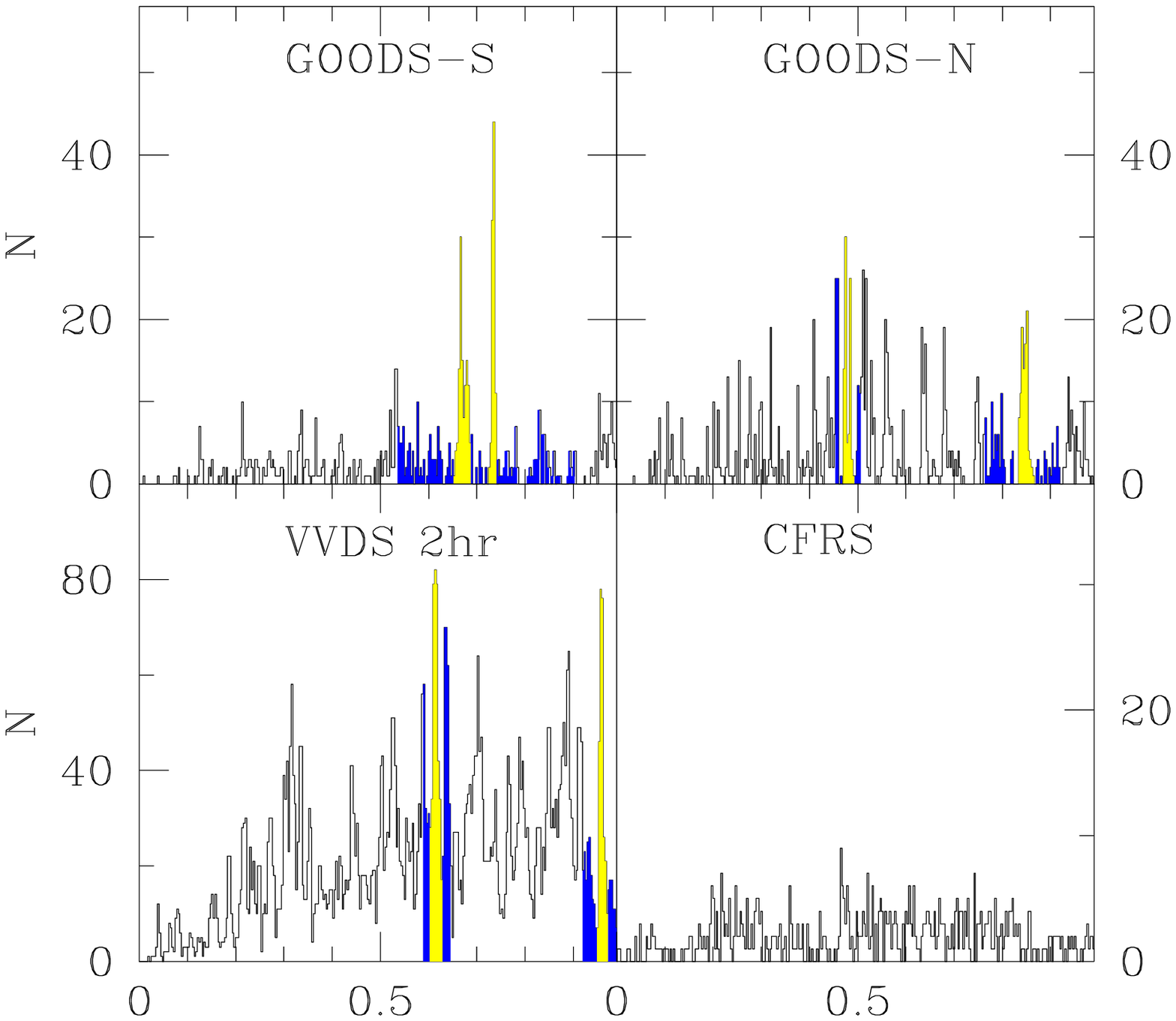}{Comparison between the redshift distribution of GOODS-S ($I_{AB} \le 23.5$, this paper)  redshift ditribution to that of GOODS-N ($R_{AB} \le 24.5$, Wirth et al, 2004), VVDS02h  ($I_{AB} \le 24$, Le Fevre et al, 2005) and CFRS ($I_{AB} \le 22.5$, Crampton et al, 2005). In yellow are shown the two most prominent structures in GOODS-S, GOODS-N and VVDS02h. In blue are identified the redshift area surrounding the structure, which contains the same number of galaxies than the two largest structures.}

\subsection{Comparison of the luminosity evolution from GOODS-S with
that from CFRS}

The CFRS sample was selected solely by the $I_{AB} \le 22.5$ criterion and includes 591 galaxies, among which 576 have $z \le 1$.
The GOODS spectroscopic catalog  contains 428 $I_{AB} \le 22.5$ galaxies, including 405 with $z \le 1$. The number of objects with $I \le 22.5$ forms about 49\%
of those present within this magnitude limit in the GOODS area and Fig.\ref{histo} (see upper middle panel) shows that it is a fair representation of  $I \le 22.5$ galaxies. In
Fig. \ref{lum} we show the redshift variation of the ratios of 
luminosity densities of observed galaxies from the GOODS catalog with that from the CFRS, for galaxies with
$0.25 \le z \le 1$, using 3 redshift bins.  Recall that luminosity densities are evaluated from integrating the luminosity functions, which are themselves estimated from the distribution of observed galaxies. Any difference between the distribution of observed galaxies in two distinct areas, would undoubtedly generate similar difference in the derived luminosity densities. Fig. \ref{lum} proves that GOODS-South field displays a strong excess of luminosity densities in the z=0.5-0.75 bin, at B, V and especially at K wavelengths.  We interpret this as related to large scale structure (i.e. cosmic variance) because more than half of the GOODS-South galaxies in that bin are found in two large scale structures (see  Fig. \ref{histo_z}). Besides this, the shape of the U luminosity density evolution is far less affected, providing that star formation rate density from UV light is almost comparable from GOODS-South to CFRS. Nevertheless, Fig. \ref{lum} suggests that one has to be cautious in interpreting evolution of the stellar mass density on the sole basis of the GOODS-South or the CDFS galaxies, since it is particularly related to the K luminosity density evolution. In the next section,
we test more accurately why the large scale
structures in GOODS-South preferentially affect red luminosity densities rather than those at blue wavelengths. 

To understand whether the cosmic variance related to the CDFS/GOODS-South field might affect analyses of galaxy evolution, Fig. \ref{lum} also displays (dashed lines) the same comparison with CFRS, but with adopting photometric redshifts rather than spectroscopic ones. Photometric redshifts are retrieved from Wolf et al. (2004) and are within the most used and accurate values existing for this field. Interestingly, most of the signature of the large scale structures disappear in the intermediate bin, except a signal in the K band. An examination of the distribution of photometric redshifts (see Wolf et al.' Figure 10) shows indeed that the large scale structures are somewhat recovered, but are much less prominent than in Fig. \ref{histo_z} (upper-left panel). For example, with photometric redshifts, the two prominent structures at z=0.670 and z=0.735 have merged, and the corresponding "equivalent width" is $\Delta$z $\sim$ 0.08, to be compared to $\Delta$z= 0.317 when using spectroscopic redshifts.  Thus, we wonder if some of the evolutionary trends found using photometric redshift surveys cannot be misunderstood as related to CDFS large scale structures.

%

\plotn{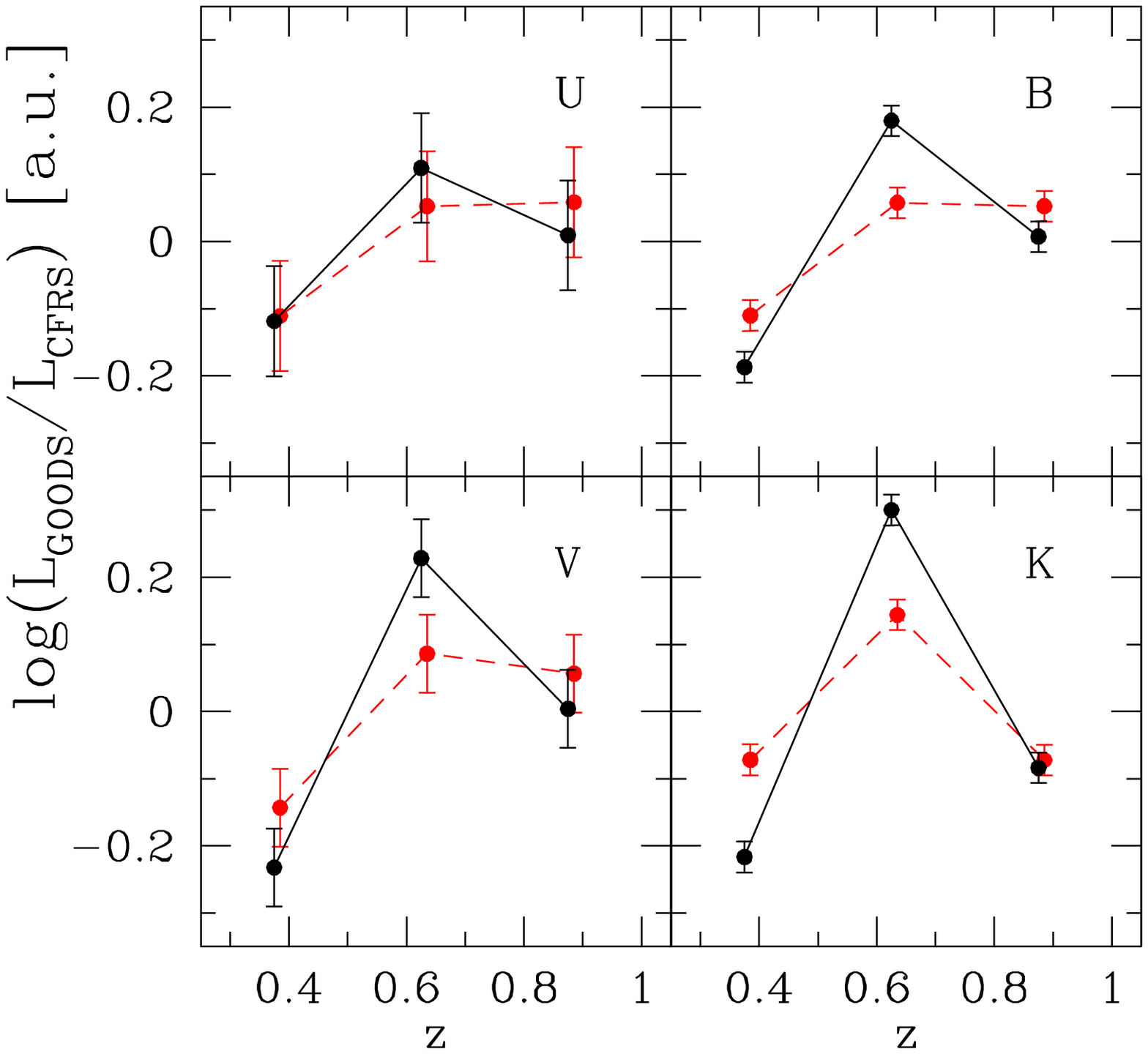}{Evolution of ratios of luminosity densities of observed galaxies
from the GOODS spectroscopic sample and observed galaxies from CFRS as a function
of redshift, for 3 redshift bins (0.25-0.5; 0.5-0.75 and 0.75-1). Calculations of absolute luminosities have been done following Hammer et al (2005). There is no noticeable variation in the evolution of U luminosity density between the samples, while the B, V and K light shows increasingly significant
variations. We find the presence of large scale structure in the GOODS
responsible for this. For consistency reason we derived the $K_{AB}$
band luminosities in CFRS using $K_{AB} = K' + 1.87 $ (as in GOODS),
instead of  $K_{AB} = K' + 1.78$ used in Lilly \etal (1995). Dashed (red) lines show a similar relation, but with adopting photometric redshifts from Wolf et al (2004) instead of spectroscopic redshifts. Error bars
correspond to Poisson statistics combined with photometric errors from measurements (see Arnouts  et al, 2001) and from calculations of the absolute magnitudes. Photometric errors at U, B, V and K wavelengths are found to be 0.07, 0.02, 0.05 and 0.02 dex median values from Arnouts et al (2001), respectively.}

\subsection{Evolution of galaxies inside and outside the structures in
the GOODS.}
It is well understood that the environments in which galaxies
reside have significant influence on their formation and evolution.
To analyze the properties of galaxies inside and outside the structures,
we designate galaxies in the field and in the structures as shown in
Fig. ~\ref{strcmp}a. The redshift range for field galaxies were selected
so that each sample contains similar numbers of galaxies while the
structures were defined solely as regions with significant excesses.
In the other subpanels of Fig. ~\ref{strcmp}, we show the distribution
of U, V, and K luminosities of galaxies inside and outside the redshift
spikes.  The two distributions in U are not significantly different,
overall suggesting a similarity in their star-formation rates. However,
for K, there is significant difference in the distributions as the
galaxies in the structures are likely to be more massive on average,
while in the case of the comparison of the V-band luminosities, there is a
difference, but it is statistically weaker.  Testing each distribution
with a Kolmogorov-Smirnov statistic confirms our visual impression,
suggesting that the significance of the similarity in each pairs of
histograms is 73, 2, and 0.6\%  for U, V, and K respectively.  This implies
that the distribution of K-band magnitudes is significantly different,
the V-band magnitude distributions are only marginally different, and
the U-band magnitude distribution are statistically the same.  It is
particularly interesting to note that the (UV traced) star formation rate
does not differ significantly for galaxies inside the structures and the
field. But the existence of what are likely to be more massive galaxies
in the structures suggests that the specific star formation rate of
galaxies in the structure are lower than that in the rest of the sample,
implying that they are more evolved.  This provides evidence that the
evolution of galaxies in and around denser environments is faster than
those in the field. In summary, the presence of large scale structures in the CDFS/GOODS-South
field might also affect studies of the specific star formation and its evolution.

\plotn{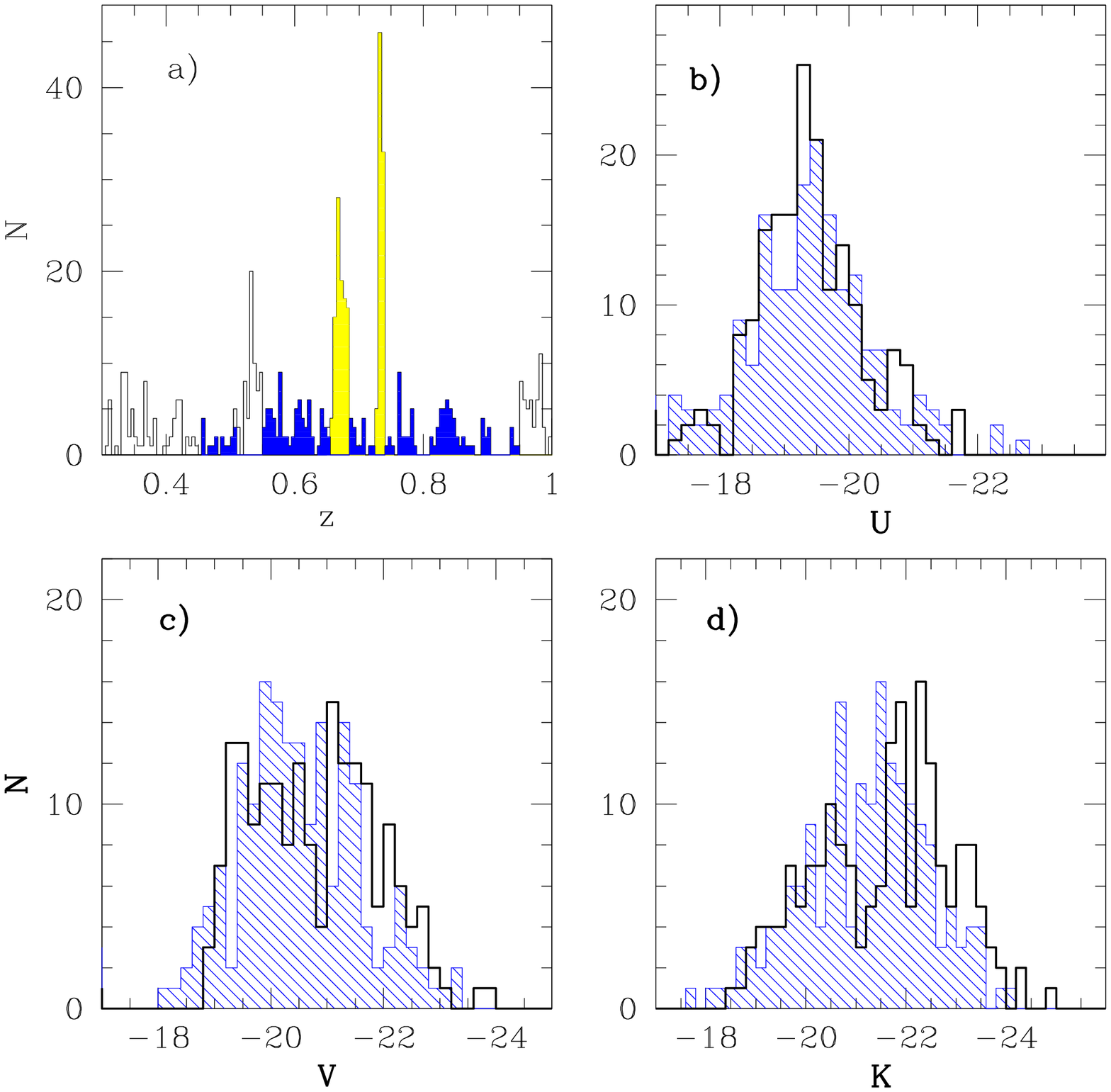}{Comparison of histograms of luminosities inside and outside
the large scale structures present in CDFS. In Fig. {\bf a}, the regions
selected to represent the structure (yellow) and the field (blue) are
shown. The redshift range for the galaxies in the two main structures is
defined by  $0.735\pm0.009$ and $0.668\pm0.016$, while the field galaxies
are taken from $0.7\pm0.25$, excluding the structures mentioned above,
and the smaller one at $0.530 \pm0.020$. In Fig.  {\bf b}, {\bf c}, and
{\bf d} we show the distributions of U, V, \and K luminosities inside
(open histogram) and outside (blue shaded) the structures.}

\section{Conclusion}

This paper prepares a series analysing the formation and
evolution of intermediate massive galaxies at intermediate redshifts.
Here we report the determination of new spectroscopic redshifts in the
Chandra Deep Field South, using low-resolution VIMOS observations. We
were able to spectroscopically identify 691 objects, consisting of
580 galaxies, 104 stars and seven QSOs. Employing an integration time
of 45 minutes for two pointings with VIMOS, this study increases  by 28\% (and 35\%) the
availability of $I_{AB} \leq 23.5$ (and $I_{AB} \leq 22.5$) galaxies  with
redshift, which were poorly represented in the existing redshift
surveys. The redshift distribution contains major peaks at
0.670, and 0.735, supporting the existence of large scale structures
already detected by the VVDS and FORS2 surveys.  The presence of large
scale structures can arise various issues related to the cosmological
variance of the GOODS.  To analyse them, we
construct a representative catalog of 640 galaxies with spectroscopic
redshift  with $I\le 23.5$ and $z\le 1$.
Hence we attempt to analyse the ``cosmological
representativeness'' of the GOODS region, by studying the properties of
galaxies in GOODS in comparison with those of other surveys, including the CFRS. It results that the GOODS-South field is by far the most affected by the presence of large scale structures.

We do find a strong excess of luminosity density at z=0.5-0.75 in GOODS-South when
compared to CFRS, especially in the V and K bands. This suggests that studies related to the evolution
of the stellar mass and/or of the specific star formation rate need to be interpreted
cautiously. By analysing the galaxies inside and outside the structures
 in the GOODS field we find that the galaxies in the structure tend to
be more massive and evolved compared to that outside the structures.
This is likely at the origin of the luminosity density excess found in GOODS-South, at this redshift. Such an effect is found to be less prominent by using photometric redshifts instead of spectroscopic redshifts. However, on the sole basis of photometric redshifts in that field, it does not seem easy to disentangle effects related to galaxy evolution from those due the presence of large scale structures, because structures appear less prominent due to the uncertainties of the redshift determination. Spectroscopic redshifts are then highly essential for galaxy evolution studies.

The GOODS South field has been observed with an unprecedented depth
at many wavelengths including optical and IR. It is one of the most
studied cosmological fields, which has generated a large number of
papers about the galaxy evolution. We stress out that many of these
studies could be affected by cosmological variance of this field as
revealed by our relatively modest spectroscopic study. We believe that
this might affect some conclusions of studies based on photometric
redshifts, especially those related to the evolution of stellar mass density and of specific star formation rate. Studies
based on spectroscopic redshifts, such as IMAGES, have also to carefully
compare the evolutionary properties of galaxies inside and outside the
structures, before concluding about an overall
picture of galaxy evolution.

\begin{acknowledgements}
CDR, AR and FH thank the Centre Franco-Indien pour la Promotion de la
Recherche Avancee (CEFIPRA) for a post-doctoral fellowship and financial
assistance respectively.  We warmly thank our referee for his/her
 detailed comments and suggestions, which have contributed a lot in improving our manuscript.

\end{acknowledgements}

\renewcommand{\baselinestretch}{1.0}
\setcounter{table}{1}
\begin{table*}
\caption {The redshift catalog of the 691 objects from the CDFS.}
{\scriptsize
\begin{tabular}{ccccrr rrrrr} \hline
ID                & T$^a$  & Q$^a$& IntID$^b$& z & I$_{iso}^c$ & M$_{U}$ & M$_{B}$ & M$_{V}$ & M$_{J}$ & M$_{K}$\\ \hline
EISJ033148.35-274816.6 &   2 &   1 & 1010003 &   0.843 &   22.66 &  -19.53 &  -20.38 &  -20.50 &   99.99 &   99.99  \\
EISJ033135.20-274713.9 &   2 &   1 & 1010004 &   0.572 &   21.83 &  -18.71 &  -19.80 &  -20.58 &   99.99 &   99.99  \\
EISJ033151.63-274823.9 &   1 &   1 & 1010006 &   0.677 &   20.81 &  -20.63 &  -21.50 &  -22.04 &   99.99 &   99.99  \\
EISJ033138.49-274721.8 &   1 &   1 & 1010007 &   0.549 &   23.10 &  -18.01 &  -18.76 &  -19.01 &   99.99 &   99.99  \\
EISJ033152.87-274824.4 &   1 &   2 & 1010009 &   0.685 &   21.69 &  -19.16 &  -20.53 &  -21.46 &   99.99 &   99.99  \\
EISJ033133.92-274641.5 &   3 &   1 & 1010013 &   1.060 &   21.47 &  -21.79 &  -22.75 &  -23.26 &   99.99 &   99.99  \\
EISJ033149.68-274751.7 &   1 &   1 & 1010014 &   0.184 &   21.55 &  -16.15 &  -17.06 &  -17.71 &   99.99 &   99.99  \\
\end{tabular}
}
\end{table*}

\setcounter{table}{1}
\begin{table*}
\caption {Cont... }
{\scriptsize
\begin{tabular}{ccccrr rrrrr} \hline
ID                & T$^a$  & Q$^a$& IntID$^b$& z & I$_{iso}^c$ & M$_{U}$ & M$_{B}$ & M$_{V}$ & M$_{J}$ & M$_{K}$ \\ \hline
EISJ033125.51-275518.1 &   2 &   1 & 1020012 &   0.744 &   21.47 &   56.86 &  -21.10 &  -22.03 &   99.99 &   99.99  \\
EISJ033130.14-275528.6 &   1 &   1 & 1020014 &   0.104 &   19.50 &  -17.22 &  -17.99 &  -18.46 &   99.99 &   99.99  \\
EISJ033145.57-275626.0 &   1 &   1 & 1020018 &   0.698 &   21.57 &  -20.14 &  -20.88 &  -21.29 &   99.99 &   99.99  \\
EISJ033130.28-275505.2 &   1 &   1 & 1020020 &   0.111 &   20.71 &  -16.11 &  -16.93 &  -17.41 &   99.99 &   99.99  \\
EISJ033148.79-275626.3 &   1 &   1 & 1020021 &   0.731 &   22.12 &  -20.14 &  -20.54 &  -20.57 &   99.99 &   99.99  \\
EISJ033119.75-275406.4 &   1 &   1 & 1020022 &   0.776 &   22.71 &  -19.08 &  -20.07 &  -20.77 &   99.99 &   99.99  \\
EISJ033151.24-275635.1 &   2 &   1 & 1020023 &   0.696 &   20.88 &  -20.40 &  -21.46 &  -22.19 &   99.99 &   99.99  \\
\end{tabular}
}
\end{table*}

\setcounter{table}{1}
\begin{table*}
\caption {Cont... }
{\scriptsize
\begin{tabular}{ccccrr rrrrr} \hline
ID                & T$^a$  & Q$^a$& IntID$^b$& z & I$_{iso}^c$ & M$_{U}$ & M$_{B}$ & M$_{V}$ & M$_{J}$ & M$_{K}$\\ \hline
EISJ033225.01-275833.5 &   2 &   2 & 1030037 &   0.538 &   21.51 &   58.54 &  -19.82 &  -20.75 &   99.99 &   99.99  \\
EISJ033228.78-275844.8 &   3 &   1 & 1030039 &   0.176 &   21.93 &  -14.69 &  -16.14 &  -17.08 &   99.99 &   99.99  \\
EISJ033206.52-275659.1 &   1 &   1 & 1030041 &   0.661 &   21.09 &  -20.00 &  -21.08 &  -21.78 &   99.99 &   99.99  \\
EISJ033235.00-275904.6 &   2 &   1 & 1030042 &   0.087 &   21.03 &  -13.80 &  -15.28 &  -16.21 &   99.99 &   99.99  \\
EISJ033225.13-275818.3 &   3 &   1 & 1030043 &   0.782 &   17.92 &  -24.17 &  -24.90 &  -25.83 &   99.99 &   99.99  \\
EISJ033209.41-275659.1 &   2 &   2 & 1030045 &   0.594 &   21.88 &  -18.86 &  -19.91 &  -20.64 &   99.99 &   99.99  \\
EISJ033206.85-275640.4 &   2 &   1 & 1030046 &   0.219 &   19.20 &  -17.74 &  -19.44 &  -20.37 &   99.99 &   99.99  \\
\end{tabular}
}
\end{table*}

\setcounter{table}{1}
\begin{table*}
\caption {Cont... }
{\scriptsize
\begin{tabular}{ccccrr rrrrr} \hline
ID                & T$^a$  & Q$^a$& IntID$^b$& z & I$_{iso}^c$ & M$_{U}$ & M$_{B}$ & M$_{V}$ & M$_{J}$ & M$_{K}$ \\ \hline
EISJ033224.62-274851.5 &   2 &   1 & 1040037 &   0.676 &   20.82 &  -20.72 &  -21.48 &  -22.03 &  -22.73 &  -22.65  \\
EISJ033220.35-274819.8 &   2 &   1 & 1040038 &   0.577 &   23.95 &  -17.87 &  -18.16 &  -18.20 &  -20.11 &  -20.02  \\
EISJ033236.20-274932.0 &   1 &   1 & 1040040 &   0.549 &   20.41 &  -19.51 &  -20.99 &  -21.92 &  -23.00 &  -22.98  \\
EISJ033227.39-274842.3 &   1 &   1 & 1040042 &   0.672 &   23.31 &  -18.74 &  -19.17 &  -19.15 &  -19.41 &  -19.03  \\
EISJ033232.74-274859.6 &   1 &   1 & 1040043 &   0.670 &   21.94 &  -19.41 &  -20.33 &  -20.89 &  -21.55 &  -21.46  \\
EISJ033221.12-274744.4 &   2 &   1 & 1040045 &   0.671 &   23.03 &  -18.82 &  -19.40 &  -19.52 &  -19.80 &  -19.64  \\
EISJ033244.72-274923.0 &   1 &   1 & 1040046 &   0.574 &   21.33 &  -19.23 &  -20.35 &  -21.07 &  -21.53 &  -21.45  \\
\end{tabular}
}
\end{table*}

\setcounter{table}{1}
\begin{table*}
\caption {Cont... }
{\scriptsize
\begin{tabular}{ccccrr rrrrr} \hline
ID                & T$^a$  & Q$^a$& IntID$^b$& z & I$_{iso}^c$ & M$_{U}$ & M$_{B}$ & M$_{V}$ & M$_{J}$ & M$_{K}$ \\ \hline
EISJ033248.70-274946.5 &   3 &   1 & 2010087 &   1.368 &   23.55 &  -21.03 &  -21.92 &  -22.42 &  -22.47 &  -22.38  \\
EISJ033251.60-275017.9 &   1 &   1 & 2010088 &   0.848 &   21.26 &  -21.26 &  -21.80 &  -21.92 &  -22.42 &  -22.27  \\
EISJ033249.68-275053.4 &   2 &   1 & 2010089 &   0.899 &   23.32 &  -19.47 &  -19.95 &  -20.04 &  -20.17 &  -19.83  \\
EISJ033222.48-274934.8 &   2 &   2 & 2010091 &   0.553 &   21.94 &  -17.30 &  -19.48 &  -20.41 &  -22.36 &  -22.34  \\
EISJ033224.74-274931.8 &   3 &   1 & 2010092 &   1.582 &   23.79 &  -21.51 &  -22.21 &  -22.67 &  -22.45 &  -22.36  \\
EISJ033223.44-275024.9 &   2 &   1 & 2010093 &   1.535 &   23.11 &  -21.89 &  -22.45 &  -22.74 &  -21.59 &  -21.23  \\
EISJ033222.61-275232.0 &   1 &   1 & 2010096 &   0.755 &   21.49 &  -20.38 &  -21.20 &  -21.61 &  -22.17 &  -22.07  \\
\end{tabular}
}
\end{table*}

\setcounter{table}{1}
\begin{table*}
\caption {Cont... }
{\scriptsize
\begin{tabular}{ccccrr rrrrr} \hline
ID                & T$^a$  & Q$^a$& IntID$^b$& z & I$_{iso}^c$ & M$_{U}$ & M$_{B}$ & M$_{V}$ & M$_{J}$ & M$_{K}$ \\ \hline
EISJ033218.87-274734.5 &   2 &   2 & 2020111 &   0.733 &   22.29 &  -18.07 &  -20.21 &  -21.14 &  -22.04 &  -22.01  \\
EISJ033233.55-274623.5 &   3 &   1 & 2020112 &   0.337 &   21.25 &  -17.28 &  -18.61 &  -19.53 &  -20.37 &  -20.28  \\
EISJ033220.49-274732.5 &   2 &   1 & 2020113 &   0.599 &   21.97 &  -17.88 &  -19.74 &  -20.67 &  -22.20 &  -22.18  \\
EISJ033209.43-274135.2 &   1 &   2 & 2020117 &   0.736 &   21.95 &  -18.40 &  -20.56 &  -21.50 &  -22.03 &  -21.96  \\
EISJ033219.39-274051.8 &   1 &   1 & 2020118 &   0.761 &   20.82 &  -20.46 &  -21.85 &  -22.77 &   99.99 &   99.99  \\
EISJ033238.27-274455.4 &   2 &   1 & 2020126 &   0.656 &   21.48 &  -18.62 &  -20.57 &  -21.50 &  -22.27 &  -22.19  \\
EISJ033238.60-274500.2 &   2 &   1 & 2020127 &   0.981 &   23.83 &  -19.28 &  -19.80 &  -19.96 &  -20.94 &  -20.84  \\
\end{tabular}
}
\end{table*}

\setcounter{table}{1}
\begin{table*}
\caption {Cont... }
{\scriptsize
\begin{tabular}{ccccrr rrrrr} \hline
ID                & T$^a$  & Q$^a$& IntID$^b$& z & I$_{iso}^c$ & M$_{U}$ & M$_{B}$ & M$_{V}$ & M$_{J}$ & M$_{K}$  \\ \hline
EISJ033153.91-275223.8 &   2 &   2 & 2040009 &   0.821 &   21.69 &  -19.26 &  -21.39 &  -22.33 &   99.99 &   99.99  \\
EISJ033200.44-275228.8 &   2 &   2 & 2040016 &   0.632 &   21.13 &  -18.72 &  -20.79 &  -21.72 &   99.99 &   99.99  \\
EISJ033138.55-275420.7 &   3 &   1 & 2040017 &   0.747 &   22.27 &  -19.84 &  -20.42 &  -20.57 &   99.99 &   99.99  \\
EISJ033149.09-275328.2 &   2 &   1 & 2040018 &   0.617 &   22.02 &  -18.59 &  -19.86 &  -20.67 &   99.99 &   99.99  \\
EISJ033137.27-275450.6 &   1 &   1 & 2040022 &   0.703 &   22.81 &  -19.19 &  -19.75 &  -19.81 &   99.99 &   99.99  \\
EISJ033155.14-275338.1 &   2 &   2 & 2040027 &   0.216 &   21.00 &  -17.77 &  -18.50 &  -18.79 &   99.99 &   99.99  \\
EISJ033143.32-275441.7 &   1 &   1 & 2040028 &   0.215 &   22.26 &  -16.77 &  -17.30 &  -17.54 &   99.99 &   99.99  \\
\end{tabular}
}
\end{table*}

\setcounter{table}{1}
\begin{table*}
\caption {Cont... }
{\scriptsize
\begin{tabular}{ccccrr rrrrr} \hline
ID                & T$^a$  & Q$^a$& IntID$^b$& z & I$_{iso}^c$ & M$_{U}$ & M$_{B}$ & M$_{V}$ & M$_{J}$ & M$_{K}$ \\ \hline
EISJ033233.56-275645.9 &   1 &   4 & 1030084 &   0.000 &   21.09 &   99.99 &   99.99 &   99.99 &   99.99 &   99.99  \\
EISJ033235.40-275625.7 &   1 &   4 & 1030090 &   0.000 &   21.14 &   99.99 &   99.99 &   99.99 &   99.99 &   99.99  \\
EISJ033210.17-275406.9 &   1 &   4 & 1030096 &   0.000 &   21.40 &   99.99 &   99.99 &   99.99 &   99.99 &   99.99  \\
EISJ033221.78-275409.9 &   1 &   4 & 1030112 &   0.000 &   19.24 &   99.99 &   99.99 &   99.99 &   99.99 &   99.99  \\
EISJ033204.00-275701.8 &   2 &   4 & 1030119 &   0.000 &   19.17 &   99.99 &   99.99 &   99.99 &   99.99 &   99.99  \\
EISJ033210.51-275547.6 &   1 &   4 & 1030121 &   0.000 &   21.62 &   99.99 &   99.99 &   99.99 &   99.99 &   99.99  \\
EISJ033217.39-275307.3 &   1 &   4 & 1030126 &   0.000 &   22.69 &   99.99 &   99.99 &   99.99 &   99.99 &   99.99  \\
\end{tabular}
  }
\begin{list}{}{}
\item[$^{\mathrm{a}}$] Classification (type=1: emission; type= 2 absorption; type= 3: QSO; type= 4: star) and associated quality of the 
spectra described in section 2 (Q= 1: insecure; Q= 2: secure; Q= 9: single emission)
\item[$^{\mathrm{b}}$] Internal identifier associated to spectra filename.
\item[$^{\mathrm{c}}$] $I_{AB}$ band photometry from EIS database (Arnouts et al., 2001)
\item[$^{\mathrm{-}}$] U, B, V, J and K absolute magnitudes in the AB system estimated following Hammer et al. (2005). 99.99 corresponds to non available quantities.

\end{list}
\end{table*}

\end{document}

%% file: bk_macros.tex
\def\be{\begin{equation}}
\def\ee{\end{equation}}
\def\bea{\begin{eqnarray}}
\def\eea{\end{eqnarray}}
\def\btable{\begin{table}}
\def\etable{\end{table}}
\def\btabular{\begin{tabular}}
\def\etabular{\end{tabular}}
\def\bdoc{\begin{document}}
\def\edoc{\end{document}}
\def\bfig{\begin{figure}}
\def\efig{\end{figure}}
\def\ben{\begin{enumerate}}
\def\een{\end{enumerate}}
\def\bit{\begin{itemize}}
\def\eit{\end{itemize}}
\def\bitem{\begin{itemize}}
\def\eitem{\end{itemize}}
\def\bquote{\begin{quote}}
\def\equote{\end{quote}}

\def\bcenter{\begin{center}}
\def\ecenter{\end{center}}

\def\bsmall{\begin{small}}
\def\esmall{\end{small}}
\def\blarge{\begin{large}}
\def\elarge{\end{large}}
\def\bLarge{\begin{Large}}
\def\eLarge{\end{Large}}
\def\bhuge{\begin{huge}}
\def\ehuge{\end{huge}}
\def\bHuge{\begin{Huge}}
\def\eHuge{\end{Huge}}
\def\bfoot{\begin{footnotesize}}
\def\efoot{\end{footnotesize}}

\def\n{\noindent}
\def\bb{\left(}
\def\eb{\right)}
\def\bbs{\left[}
\def\ebs{\right]}
\def\bba{\left<}
\def\eba{\right>}
\def\bbsq{\left[}
\def\ebsq{\right]}
\def\bbcu{\left\{}
\def\ebcu{\right\}}
\def\bml{\left|}
\def\emr{\right|}
\newcommand \sib [1] {\bb #1 \eb}

\newcommand \sqb [1] {\bbsq #1 \ebsq}
\newcommand \cub [1] {\bbcu #1 \ebcu}
\def\etal{et al.\ }
\def\ie{i.\,e.\,, }
\def\eg{e.\,g.\,} 
\def\etc{etc.\ }
\def\viz{viz.\,, }

\def\gne #1#2{\ \vphantom{S}^{\raise-0.5pt\hbox{$\scriptstyle#1$}}_
{\raise0.5pt \hbox{$\scriptstyle#2$}}}
\def\gneq{\gne > \sim}
\def\lneq{\gne < \sim}

\def\mod #1{|#1|}
\def\vec#1{{\bf #1}}
\def\dotvec#1{\dot\vec{#1}}
\def\mvec#1{\mod{\vec{#1}}}
\def\mean #1{\bba #1 \eba}
\def\meanvsq#1{\mean{\vec{#1}^2}}
\def\dotp#1#2{\vec{#1}\cdot\vec{#2}}
\def\meandotp#1#2{\mean{\dotp{#1}{#2}}}
\def\crossp#1#2{\vec{#1}\times\vec{#2}}
\def\vecp{\vec{p}}
\def\vecpone{\vec{p}_{1}}
\def\vecn{\vec{n}}
\def\vecnone{\vec{n}_{1}}

\def\reci #1{\frac {1} {#1}}
\def\ten#1#2{#1{\times10^{#2}}}
\def\onlyten#1{10^{#1}}
\def\deg{\unit{deg}}
\def\degsq{\unit{deg^2}}
\def\ster{\unit{steradian}}
\def\perc{\unit{percent}}
\def\percent{\unit{percent}}

\def\deri#1#2{\frac {d#1} {d#2}}
\def\deriemp#1{\frac {d} {d#1}}
\def\pderi#1#2{\frac {\partial#1} {\partial#2}}
\def\pderiemp#1{\frac {\partial} {\partial#1}}
\def\secderi#1#2{\frac {d^2#1} {d#2^2}}
\def\secpderi#1#2{\frac {\partial^2#1} {\partial#2^2}}

\def\grad{\nabla}

\def\unit #1{\,{\rm #1}} 

\def\angst{\unit{\AA}}

\def\gm{\unit{gm}}
\def\gmi{\gm^{-1}}
\def\mg{\unit{mg}}
\def\kg{\unit{kg}}
\def\nm{\unit{nm}}
\def\mum{\unit{\mu m}}
\def\mumi{\unit{\mu\,m^{-1}}}
\def\mm{\unit{mm}}
\def\cm{\unit{cm}}
\def\mtr{\unit{m}}
\def\mtri{\unit{m^{-1}}}
\def\km{\unit{km}}

\def\pc{\unit{pc}}
\def\kpc{\unit{kpc}}
\def\mpc{\unit{Mpc}}
\def\gpc{\unit{Gpc}}

\def\cmi{\unit{cm^{-1}}}
\def\cmsq{\unit{cm^2}}
\def\cmsqi{\unit{cm^{-2}}}
\def\cmcu{\unit{cm^3}}
\def\cmcui{\unit{cm^{-3}}}
\def\kmi{\unit{km^{-1}}}
\def\pci{\unit{pc^{-1}}}
\def\kpci{\unit{kpc^{-1}}}
\def\mtrsq{\unit{m^2}}
\def\mtrsqi{\unit{m^{-2}}}
\def\mtrcu{\unit{m^3}}
\def\mtrcui{\unit{m^{-3}}}
\def\mpccu{\unit{cm^3}}
\def\mpccui{\unit{cm^{-3}}}
\def\gpccu{\unit{cm^3}}
\def\gpccui{\unit{Gpc^{-3}}}
\def\kpsi{\unit{km \, s^{-1}}}

\def\ksec{\unit{ksec}}
\def\msec{\unit{msec}}
\def\musec{\unit{\mu\,sec}}
\def\min{\unit{min}}
\def\hr{\unit{hr}}
\def\day{\unit{day}}
\def\days{\unit{days}}
\def\week{\unit{week}}
\def\yr{\unit{yr}}
\def\myr{\unit{Myr}}
\def\gyr{\unit{Gyr}}
\def\micron{\unit{\mu m}}
\def\delto{\Delta t_{\rm o}}

\def\seci{\unit{s^{-1}}}
\def\dayi{\unit{day^{-1}}}
\def\yri{\unit{yr^{-1}}}

\def\cms{\unit{cm\,s^{-1}}}
\def\mts{\unit{m\,s^{-1}}}
\def\mtss{\unit{m\,s^{-2}}}
\def\kms{\unit{km\,s^{-1}}}
\def\kmh{\unit{km\,h^{-1}}}

\def\mas{\unit{mas}}
\def\muas{\unit{{\mu}}as}
\def\masyr{\unit{mas\,yr^{-1}}}
\def\arcs{\unit{arcsec}}
\def\arcsec{\unit{arcsec}}
\def\arcsecsq{\unit{arcsec^2}}
\def\arcsecsqi{\unit{arcsec^{-2}}}
\def\arcmin{\unit{arcmin}}
\def\arcm{\unit{arcmin}}
\def\arcmsq{\unit{arcmin^2}}
\def\degsq{\unit{deg^2}}
\def\degsqi{\unit{deg^{-2}}}

\def\vlos{\vec{v}_{\rm los}}

\def\hz{\unit{Hz}}
\def\khz{\unit{kHz}}
\def\mhz{\unit{MHz}}
\def\ghz{\unit{GHz}}
\def\thz{\unit{Thz}}
\def\ev{\unit{eV}}
\def\kev{\unit{keV}}
\def\mev{\unit{MeV }}
\def\gev{\unit{GeV }}
\def\tev{\unit{TeV }}

\def\kel{\unit{K}}

\def\newt{\unit{N}}

\def\funit{\rm\,erg\,cm^{-2}\,sec^{-1}}
\def\pfunit{\unit{cm^{-2}\,sec^{-1}}}
\def\funith{\rm\,erg\,cm^{-2}\,sec^{-1}\,Hz^{-1}}
\def\funitb{\rm\,erg\,cm^{-2}\,sec^{-1}\ster^{-1}}
\def\funitbh{\rm\,erg\,cm^{-2}\,sec^{-1}\Hz^{-1}\ster^{-1}}
\def\funitba{\rm\,erg\,cm^{-2}\,sec^{-1}\AA^{-1}\ster^{-1}}
\def\punitb{\rm\,photons\,cm^{-2}\,sec^{-1}\ster^{-1}}
\def\lunit{\rm\,erg\,sec^{-1}}
\def\lunith{\rm\,erg\,sec^{-1}\,Hz^{-1}}
\def\whz{\unit{W\,Hz^{-1}}}
\def\whzs{\unit{W\,Hz^{-1}\,str^{-1}}}
\def\wmtrsqi{\unit{W\,m^{-2}}}
\def\jy{\unit{Jy}}
\def\mjy{\unit{mJy}}
\def\mujy{\unit{\mu Jy}}
\def\jyu{\unit{W\,m^{-2}\,Hz}}
\def\erg{\unit{erg}}
\def\ergs{\unit{erg\,s^{-1}}}
\def\ergsh{\unit{erg\,s^{-1}\,Hz^{-1}}}
\def\mcrab{\unit{milliCrab}}
\def\ergsc{\unit{erg\,s^{-1}\,cm^{-3}}}
\def\ergcmcu{\unit{erg\,cm^{-3}}}
\def\ergscmcu{\unit{erg\,s^{-1}\,cm^{-3}}} 
\def\ergcmt{\unit{erg\,cm^{-3}}}
\def\ergcmcui{\unit{erg\,cm^{-3}}}

\def\gmcmcui{\gm\cmcui}
\def\kgmtrcui{\kg\mtrcui}
 
\def\melec{m_{\rm e}}
\def\mectwo{\melec c^2}
\def\mprot{m_{\rm p}}
\def\nelec{n_{\rm e}}
\def\nion{n_{\rm i}}
\def\nprot{n_{\rm p}}
\def\pelec{p_{\rm e}}
\def\pfermi{p_{\rm f}}
\def\efermi{E_{\rm f}}
\def\mue{\mu_{\rm e}}

\def\gntff{\overline{g}_{\rm ff}}

\def\eneutrino{\nu_{\rm e}}
\def\eaneutrino{\overline{\nu}_{\rm e}}
\def\mneutrino{\nu_{\mu}}
\def\maneutrino{\\overline{nu}_{\mu}}
\def\tneutrino{\nu_{\tau}}
\def\maneutrino{\\overline{nu}_{\tau}}

\def\fine{\alpha_{\rm f}}
\def\comptw{\lambda_{\rm e}}

\def\threehe{^3{\rm He}}
\def\fourhe{^4{\rm He}}
\def\sixli{^6{\rm Li}}
\def\sevenli{^7{\rm Li}}
\def\ninebe{^9{\rm Be}}
\def\tenb{^{10}{\rm B}}
\def\elevenb{^{11}{\rm B}}
\def\twelvec{^{12}{\rm C}}
\def\sixteeno{^{16}{\rm O}}

\def\fourhebyh{\fourhe/{\rm H}}
\def\threehebyh{\threehe/{\rm H}}
\def\sevenlibyh{^7{\rm Li}/{\rm H}}
\def\bsqfh{{\rm B^2FH}}

\def\mag{\unit{magnitude}}
\def\mags{\unit{magnitudes}}
\def\absm#1{M_{\rm#1}}
\def\appm#1{m_{\rm#1}}
\def\Delm#1{\Delta m_{#1}}
\def\magarcsecsqi{\unit{mag\arcsecsqi}}

\def\lsol{L_{\odot}}
\def\msol{M_{\odot}}
\def\rsol{R_{\odot}}
\def\zsol{Z_{\odot}}

\def\aunit{\unit{AU}}

\def\mear{M_{\oplus}}
\def\rear{R_{\oplus}}
\def\aear{a_{\oplus}}
\def\pear{P_{\oplus}}

\def\mjup{M_{\rm Jupiter}}
\def\ajup{a_{\rm Jupiter}}
\def\pjup{P_{\rm Jupiter}}
\def\msat{M_{\rm Saturn}}

\def\amer{a_{\rm Mercury}}
\def\pmer{P_{\rm Mercury}}

\def\aplu{a_{\rm Pluto}}
\def\pplu{P_{\rm Pluto}}

\def\meight{\bb\frac{M}{\onlyten{8}\msol}\eb}

\def\mbymsol{\bb\frac{M}{\msol}\eb}

\def\aop{\alpha_{\rm op}}
\def\ar{\alpha_{\rm R}}
\def\ax{\alpha_{\rm x}}
\def\ag{\alpha_{\gamma}}
\def\aire{\alpha_{2.2-25\micron}}
\def\asm{\alpha_{\rm sm}}
\def\onemal{1-\alpha}
\def\nuone{\nu_1}
\def\nutwo{\nu_2}
\def\nuz{\nu_0}
\def\nug{\nu_g}
\def\nuc{\nu_c}
\def\nua{\nu_a}
\def\nup{\nu_p}
\def\nux{\nu_{\rm x}}
\def\nuop{\nu_{\rm op}}
\def\nupz{\nu_{p0}}
\def\nuaop{\nu^{-\alpha_{op}}}
\def\nualpha{\nu^{-\alpha}}
\def\alnu{\alpha_{\nu}}
\def\alnua{\alpha_{\nu_a}}
\def\alnusc{\alnu^{\rm sc}}
\def\jnu{j_{\nu}}
\def\jnua{j_{\nu_a}}
\def\tstar{\tau_{*}}
\def\tnu{\tau_{\nu}}
\def\etnu{e^{\tnu}}
\def\emtnu{e^{-\tnu}}
\def\fnu{F(\nu)}
\def\fnua{F(\nua)}
\def\fr{F_{\rm R}}
\def\fop{F_{\rm op}}
\def\fopl{F_{\rm op}^l}
\def\fx{F_{\rm x}}
\def\fxl{F_{\rm x}^{l}}
\def\frlim{F_{R,lim}}
\def\mv{m_V}
\def\mb{m_B}
\def\mr{m_R}
\def\Mv{M_V}
\def\Mb{M_B}
\def\Mr{M_R}

\def\bnu{B(\nu)}
\def\blnu{B_{\nu}}
\def\inu{I(\nu)}
\def\ilnu{I_{\nu}}
\def\snu{S(\nu)}
\def\slnu{S_{\nu}}
\def\Snu{S(\nu)}
\def\unu{u(\nu)}
\def\ulnu{u_{\nu}}
\def\Snua{S(\nu_a)}
\def\nnu{n(\nu)}
\def\nlnu{n_{\nu}}
\def\nomega{n(\omega)}
\def\nlomega{n_{\omega}}
\def\taunu{\tau_{\nu}}
\def\etaunu{e^{\taunu}}
\def\emtaunu{e^{-\taunu}}
\def\exptaunu{\exp(\taunu)}
\def\expmtaunu{\exp(-\taunu)}
\def\planckinu{\frac{2h\nu^3/c^2}{\exp(h\nu/kT)-1}}
\def\planckunu{\frac{8\pi h\nu^3/c^3}{\exp(h\nu/kT)-1}}
\def\planckilambda{\frac{2hc^2/\lambda^5}{\exp(hc/\lambda kT)-1}}
\def\planckulambda{\frac{8\pi hc/\lambda^5}{\exp(hc/\lambda kT)-1}} 

\def\plisnu{\frac{2h\nu^3/c^2}{\exp(h\nu/kT)-1}}
\def\plusnu{\frac{8\pi h\nu^3/c^3}{\exp(h\nu/kT)-1}}
\def\plislambda{\frac{2hc^2/\lambda^5}{\exp(hc/\lambda kT)-1}}
\def\pluslambda{\frac{8\pi hc/\lambda^5}{\exp(hc/\lambda kT)-1}} 
\def\plnut{\frac{2h\nu^3/c^2}{\exp(h\nu/kT)-1}}
\def\rjnut{\frac{2kT\nu^2}{c^2}}

\def\Pnu{P(\nu)}
\def\lnu{L(\nu)}
\def\llnu{\log \lnu}
\def\lr{L_{\rm R}}
\def\llr{\log L_{\rm R}}
\def\lrlim{L_{R,lim}}
\def\lfive{L(5\ghz)}
\def\lop{L_{{\rm op}}}
\def\llop{\log L_{{\rm op}}}
\def\lx{L_{\rm x}}
\def\llx{\log L_{\rm x}}
\def\lxmin{L_{\rm x,min}}
\def\lxmax{L_{\rm x,max}}
\def\lmax{L_{|rm max}}
\def\lir{L_{\rm IR}}
\def\luv{L_{\rm UV}} 
\def\lrc{L_{\rm rc}}
\def\llrc{\log L_{\rm rc}}
\def\lrext{L_{\rm \rm r,ext}}
\def\llrext{\log L_{r,ext}}
\def\lre{L_{\rm re}}
\def\llre{\log L_{\rm re}}
\def\lxb{L_{\rm xb}}
\def\llxb{\log L_{\rm xb}}
\def\lxu{L_{\rm xu}}
\def\llxu{\log L_{\rm xu}}
\def\lre{L_{\rm re}}
\def\llre{\log L_{\rm re}}
\def\rt{R_{\rm t}}
\def\rx{R_{\rm x}}
\def\rxbu{R_{\rm xbu}}
\def\rtx{R_{\rm tx}}
\def\gxbp{g_{\rm x}(\beta, \psi)}
\def\grbp{g_{\rm r}(\beta, \psi)}
\def\aox{\alpha_{\rm ox}}
\def\aoxl{\alpha_{\rm ox}^l}
\def\aro{\alpha_{\rm ro}}
\def\arx{\alpha_{\rm rx}}
\def\aoxx{\alpha_{\rm oxx}}
\def\auv{\alpha_{\rm uv}}
\def\sigt{\sigma_{\rm T}}
\def\sigkn{\sigma_{\rm KN}}
\def\signu{\sigma_{\nu}}
\def\siggg{\sigma_{\gamma\gamma}}
\def\tausc{\tau_{\rm sc}}
\def\taut{\tau_{\rm T}}
\def\nsc{N_{\rm sc}}
\def\freenu{l_{\nu}}
\def\U #1{U_{\rm #1}}
\def\eps{\epsilon}
\def\epsone{\epsilon_1}
\def\epstwo{\epsilon_2}
\def\epsx{\epsilon_{\rm x}}
\def\epsxs{\eps_{\rm xs}}
\def\epss{\epsilon_{\rm s}}
\def\epsg{\epsilon_{\gamma}}
\def\epsnu{\epsilon_{\nu}}
\def\epsir{\eps_{\rm IR}}
\def\epsi{\eps_{\rm i}}
\def\epsf{\eps_{\rm f}}
\def\DOM{\Delta\Omega}
\def\omegaone{\omega_{1}}

\def\hnu{h\nu}
\def\hnubyc{\frac{h\nu}{c}}
\def\hbarom{\hbar\omega}

\def\lcapc{L_{\rm C}}
\def\lcaps{L_{\rm S}}
\def\lcapi{L{\rm i}}
\def\lcaph{L_{\rm h}}
\def\dtvar{\Delta t_{\rm var}}
\def\tbri{T_{\rm b}}
\def\tsri{T_{\rm i}}
\def\Tvar{T_{\rm var}}
\def\dmv{dm_{\rm V}}
\def\betaa{\beta_{\rm a}}
\def\vela{v_{\rm a}}
\def\te{t_{\rm e}}
\def\dte{dt_{\rm e}}
\def\gammap{\gamma_{\rm p}}
\def\gammab{\gamma_{\rm b}}
\def\gammai{\gamma{\rm i}}
\def\gammaj{\gamma{\rm j}}
\def\gammac{\gamma{\rm c}}

\def\vbyvm{\frac {V} {V_m}}
\def\vbyvmp{\frac {V'} {V_m'}}
\def\mvbyvm{\bba\vbyvm\eba}
\def\mvbyvmp{\bba\vbyvmp\eba}

\def\vbyvml{V/V_m}
\def\vbyvmpl{V'/V_m'}
\def\mvbyvml{\bba\vbyvml\eba}
\def\mvbyvmpl{\bba\vbyvmpl\eba}

\def\bperp{B_{\perp}}
\def\bpar{B_{\parallel}}
\def\bperppar{B_{\perp-\parallel}}
\def\dtvar{\Delta t_{var}}
\def\Tvar{T_{var}}
\def\dmv{dm_v}
\def\betaa{\beta_a}
\def\gammap{\gamma_{\rm p}}
\def\gammab{\gamma_{\rm b}}
\def\gammamax{\gamma_{\rm max}}
\def\scrid{{\cal D}}
\def\scril{{\cal L}}
\def\scrio{{\cal O}}
\def\scrir{{\cal R}}
\def\fbeam{F_b}
\def\fext{F_{ext}}
\def\fin{F_{\rm in}}
\def\fout{F_{\rm out}}
\def\bcospsi{\beta\cos\psi}
\def\bsinpsi{\beta\sin\psi}
\def\rnu{R_{\nu}}
\def\rnuone{R_{\nu_1}}
\def\rnutwo{R_{\nu_2}}

\def\nh{N_{\rm H}}
\def\mh{M_{\rm H}}
\def\eax{E^{-\ax}}
\def\ex{E_{\rm x}}
\def\ex{E_{\rm xs}}

\def\zmin{z_{\rm min}}
\def\zmax{z_{\rm max}}
\def\rmin{r_{\rm min}}
\def\rmax{r_{\rm max}}
\def\rsc{r_{\rm sc}}
\def\rscm{r_{\rm sc,m}}

\def\corrxz{r_{\rm x,z}}
\def\corrxop{r_{\rm x,op}}
\def\corropz{r_{\rm op,z}}
\def\corrxopz{r_{\rm x,op;z}}
\def\chisq{\chi^2}
\def\chisqr{\chi^{2}_{\nu}}

\def\philx{\Phi{(\lx)}}
\def\phixstar{\Phi_{X}^{*}}
\def\phixstaro{\Phi_{X1}^{*}}
\def\lxstar{\lx^{*}}
\def\lxfour{L_{{\rm x},44}}
\def\lxstarfour{L_{{\rm x},44}^*(0)}

\def\asca{{\em ASCA }}
\def\axaf{{\em AXAF }}
\def\ein{{\em EINSTEIN }}
\def\exosat{{\em EXOSAT }}
\def\ginga{{\em GINGA }}
\def\rosat{{\em ROSAT }}
\def\compton{{\em COMPTON }}
\def\granat{{\em GRANAT }}
\def\iras{{\em IRAS }}
\def\iue{{\em IUE }}

\def\seyone{Seyfert\,1 }
\def\seytwo{Seyfert\,2 }
\def\bllac{BL Lac }
\def\bllacs{BL Lacs }
\def\rbl{{\rm RBL }}
\def\xbl{{\rm XBL }}
\def\ipc{{\rm IPC }}
\def\pspc{{\rm PSPC }}

\def\cfour{{\rm C\,IV }}
\def\cthreesf{{\rm C\,III] }}
\def\fetwo{{\rm Fe\,II }}
\def\halpha{{\rm H\,\alpha }}
\def\hbeta{{\rm H\,\beta }}
\def\hgamma{{\rm H\,\gamma}}
\def\hi{{\rm H\,I }}
\def\lyal{{\rm Ly\,\alpha }}
\def\mgtwo{{\rm Mg\,II }}
\def\ntwo{{\rm [N\,II]}}
\def\nfive{{\rm N\,V }}
\def\otwo{{\rm [O\,II] }}
\def\othree{{\rm [O\,III] }}
\def\kalpha{K_{\alpha} }
\def\ykz{Y^{K}_{Z} }

\def\const{{\rm constant}}
\def\gray{\gamma-{\rm ray}}
\def\grays{\gamma-{\rm rays}}
\def\maxtau{\tausc(1+\tausc)}
\def\maxtnu{\tnu(1+\tnu)}
\def\dm{{\rm DM}}
\def\dmsinmb{\dm\sin\mod{b}}
\def\dmsinb{\dm\sin b}
\def\sbyn{\frac{S}{N}}

\def\noi{\noindent}
\def\vf{\vspace{0.5cm}}
\def\vff{\vspace{1cm}}
\def\pbr{\pagebreak}
\def\pnew{\newpage}
\def\pemp{\pagestyle{empty}}
\def\tpemp{\thispagestyle{empty}}

\newcommand{\onlytitle}[1]
{\title{#1}
\author{ }
\date{ }
\maketitle\tpemp }

\newcommand{\onlyfig}[2]
{\begin{center}\epsfig{figure=#1.ps, width=#2in}\end{center}}

\newcommand{\onlyfige}[2]
{\begin{center}\epsfig{figure=#1.eps, width=#2in}\end{center}}

\def\vbyc{\frac{v}{c}}
\def\vbycsq{\frac{v^2}{c^2}}
\def\omvbycsq{\sqrt{1-\vbycsq}}

\def\gmr{\frac{GM}{r}}
\def\gmcsr{\frac{GM}{c^2r}}
\def\gmcsrl{GM/c^2r}
\def\tgmcsr{\frac{2GM}{c^2r}}
\def\schwarz{r_{\rm S}}
\def\tschwarz{t_{\rm S}}
\def\rergo{r_{\rm E}}
\def\rplus{r_{+}}
\def\mbh{M_{\rm BH}}
\def\mbyl{M/L}

\def\lh{l_{\rm h}}
\def\li{l_{\rm i}}
\def\ls{l_{\rm s}}

\def\omk{\Omega_{\rm K}(r)}
\def\velr{v_{r}}
\def\velp{v_{\phi}}
\def\velz{v_{z}}
\def\sigrt{\Sigma(r,t)}
\def\mach{{\cal M}}
\def\scrir{{\cal R}}
\def\tin{t_{\rm in}}
\def\tl{t_{\rm L}}
\def\ldisk{L_{\rm disk}}
\def\temps{T_{\rm s}}
\def\tempc{T_{\rm c}}
\def\tempb{T_{\rm B}}
\def\bnu{B_{\nu}}
\def\bnuts{B_{\nu}(\temps)}
\def\bnutsr{B_{\nu}(\temps(r))}

\def\rout{r_{\rm out}}
\def\tvis{t_{\rm vis}}
\def\pgas{p_{\rm g}}
\def\prad{p_{\rm r}}
\def\csou{c_{\rm s}}
\def\rstar{r_{*}}
\def\ledd{L_{\rm Edd}}
\def\mdedd{\dot{M}_{\rm Edd}}
\def\tedd{t_{\rm Edd}}
\def\tempedd{T_{\rm Edd}}
\def\bedd{B_{\rm Edd}}
\def\mdot{{\dot M}}
\def\geff{\vec{g}_{\rm eff}}
\def\phir{\Phi_{\rm rot}}
\def\phie{\Phi_{\rm eff}}
\def\zsr{z_{\rm s}(r)}
\def\udzero{u_0}
\def\udphi{u_{\phi}}
\def\uuzero{u^0}
\def\uuphi{u^{\phi}}

\def\isnu{I_{\nu}}
\def\islambda{I_{\lambda}}
\def\fsnu{F_{\nu}}
\def\fslambda{F_{\lambda}}

\def\plisnu{\frac{2h\nu^3/c^2}{\exp(h\nu/kT)-1}}
\def\plusnu{\frac{8\pi h\nu^3/c^3}{\exp(h\nu/kT)-1}}
\def\plislambda{\frac{2hc^2/\lambda^5}{\exp(hc/\lambda kT)-1}}
\def\pluslambda{\frac{8\pi hc/\lambda^5}{\exp(hc/\lambda kT)-1}} 

\def\isnurj{\isnu^{\rm RJ}}
\def\isnuw{\isnu^{\rm W}}

\def\cv{C_V}
\def\cp{C_p}
\def\dqdtv{\bb\pderi{Q}{T}\eb_V}
\def\dqdtp{\bb\pderi{Q}{T}\eb_p}

\def\CH{{\it CH}}

\def\hub{H_0}
\def\hubi{H^{-1}_0}
\def\hubble#1{H_0={#1}\unit{km\,sec^{-1}\,Mpc^{-1}}}
\def\hubbleunit{\unit{km\,sec^{-1}\,Mpc^{-1}}}
\def\hubhund{h_{100}}
\def\hubhundcu{\hubhund^3}
\def\hubhundcui{\hubhund^{-3}}
\def\hubhundi{\hubhund^{-1}}
\def\hubhundis{\hubhund^{-2}}
\def\cuponh{\bb \frac {c} {H_0} \eb}

\def\qz{q_0}
\def\qzsq{\qz^2}
\def\fqz{\bbcu\reci{\qzsq}\bbsq \qz z + (q_0-1)(\sqrt{1+2q_0z}-1)
\ebsq\ebcu}
\def\fqzno{\qz z + (q_0-1)(\sqrt{1+2q_0z}-1)}
\def\omb{\Omega_{{\rm b}}}
\def\omhi{\Omega_{{\rm H\,I}}}
\def\omlls{\Omega_{{\rm LLS}}}
\def\omlb{\Omega_{{\rm lb}}}
\def\omigm{\Omega_{{\rm IGM}}}
\def\omdlyal{\Omega_{{\rm DLy\,\alpha}}}
\def\zpr{z^{\prime}}
\def\zabs{z_{{\rm abs}}}
\def\zem{z_{{\rm em}}}
\def\zlim{z_{\rm lim}}
\def\zlimg{z_{\rm lim,G}}
\def\zlimgq{z_{\rm lim,GQ}}
\def\dlum{D_{\rm L}}
\def\dlumsq{D^2_{\rm L}}
\def\dmet{d_{\rm m}}
\def\dang{d_{\rm A}}

\def\lamo#1{\lambda_{\rm o_{#1}}}
\def\lame#1{\lambda_{\rm e_{#1}}}

\def\re{r_{\rm e}}
\def\rbyre{\frac{r}{\re}}
\def\rbyredev{\bb\rbyre\eb^{1/4}}
\def\rbyren{\bb\rbyre\eb^{1/n}}
\def\rbyrel{r/\re}
\def\rbyredevl{(\rbyrel)^{1/4}}
\def\rbyrenl{(\rbyrel)^{1/n}}
\def\rebyrd{\frac{\re}{\rd}}
\def\rebyrdl{\re/\rd}

\def\izero{I(0)}
\def\ire{I(\re)}
\def\ir{I(r)}
\def\icr{I_{\rm c}(r)}

\def\muzero{\mu(0)}
\def\mure{\mu(\re)}
\def\mur{\mu(r)}

\def\ibzero{I_{\rm b}(0)}
\def\ibre{I_{\rm b}(\re)}
\def\mibre{\mean{\ibre}}
\def\iblre{I_{\rm b}(<\re)}
\def\ibr{I_{\rm b}(r)}
\def\iblr{I_{\rm b}(<r)}
\def\ibtot{I_{\rm b,tot}}
\def\ibxy{I_{\rm b}(x,y)}
\def\isky{I_{\rm sky}}

\def\mubzero{\mu_{\rm b}(0)}
\def\mubre{\mu_{\rm b}(<\re)}
\def\mmubre{\mean{\mubre}}
\def\mubr{\mu_{\rm b}(r)}
\def\mulim{\mu_{\rm lim}}

\def\mudzero{\mu_{\rm d}(0)}
\def\mugzero{\mu_{\rm G}(0)}

\def\rd{r_{\rm d}}
\def\rs{r_{\rm s}}
\def\rbyrd{\frac{r}{\rd}}
\def\rbyrdl{r/\rd}
\def\idzero{I_{\rm d}(0)}
\def\idr{I_{\rm d}(r)}
\def\idlr{I_{\rm d}(<r)}
\def\idtot{I_{\rm d,tot}}
\def\idxy{I_{\rm d}(x,y)}

\def\imax{I_{\rm max}}
\def\imin{I_{\rm min}}

\def\dev{\onlyten{-3.33\rbyredevl}}
\def\devn{\onlyten{-\cn\rbyren}}
\def\exprbyrd{e^{-(\rbyrdl)}}
\def\cn{c_n}

\def\dbyb{\frac{D}{B}}
\def\bbyd{\frac{B}{D}}
\def\dbybl{D/B}
\def\bbydl{B/D}

\def\absgal{M_{\rm G}}
\def\absbul{M_{\rm B}}
\def\absdisk{M_{\rm d}}
\def\absq{M_{\rm Q}}

\def\lstar{L_{*}}

\def\mbyl{\bb\frac{M}{L}\eb}

\def\labchap #1{\label{chap:#1}}
\def\labequn #1{\label{eq:#1}}
\def\labfig #1{\label{fig:#1}}
\def\labsecn #1{\label{sec:#1}}
\def\labsubsecn #1{\label{subsecn:#1}}
\def\labsubsubsecn #1{\label{subsubsecn:#1}}
\def\labtablem #1{\label{tab:#1}}
\def\labapp #1{\label{app:#1}}
\def\labboxl #1{\label{box:#1}}

\def\boxl #1{box~\ref{box:#1}}
\def\chap #1{Chapter~\ref{chap:#1}}
\def\equn #1{Equation~\ref{eq:#1}}
\def\fig #1{Figure~\ref{fig:#1}}
\def\relation #1{relation~\ref{eq:#1}}
\def\secn #1{Section~\ref{sec:#1}}
\def\subsecn #1{Section~\ref{subsecn:#1}}
\def\subsubsecn #1{Section~\ref{subsubsecn:#1}}
\def\app#1{Appendix~\ref{app:#1}}
\def\tablem #1{Table~\ref{tab:#1}}

\def\dequn#1#2{Equations~{\ref{eq:#1}}~and~{\ref{eq:#2}}}
\def\tequn#1#2#3{Equations~{\ref{eq:#1}},~{\ref{eq:#2}} and ~{\ref{eq:#3}}}
\def\mequn#1#2{Equations~{\ref{eq:#1}}~to~{\ref{eq:#2}}}
\def\dfig#1#2{Figures~{\ref{fig:#1}}~and~{\ref{fig:#2}}}
\def\mfig#1#2{Figures~{\ref{fig:#1}}~to~{\ref{fig:#2}}}
\def\dtablem#1#2{Tables~{\ref{tab:#1}}~and~{\ref{tab:#2}}}
\def\dsubsecn#1#2{Subsections~{\ref{subsecn:#1}}~and~{\ref{subsecn:#2}}}

\def\tobs#1{{\it The Observatory}{\bf #1}}

\def\cfa#1{{\it Center for Astrophysics Preprint }{\bf #1}}
\def\asp{Astron. Soc. of the Pacific, San Francisco.}
\def\clar{Clarendon Press, Oxford.} 
\def\cup{Cambridge University Press, Cambridge.}
\def\free{W. H. Freeman, New York.}
\def\klu{Kluwer, Dordrecht.}
\def\nrao{NRAO, Greenbank.}
\def\plenumn{PLenum Press, New York.}
\def\prince{Princeton University Press, Princeston}
\def\ridel{Ridel, Dordrecht.}
\def\springerb{Springer-Verlag, Berlin.}
\def\springern{Springer-Verlag, New York.}
\def\wiley{Wiley, New York.}
\def\yale{Yale University Press, New Haven.}

\def\noin{\noindent}

\def\sqr#1#2{{\vcenter{\vbox{\hrule height.#2pt
  \hbox {\vrule width.#2pt height#1pt \kern#1pt
  \vrule width.#2pt}
  \hrule height.#2pt}}}}

\def\square{\mathchoice\sqr68\sqr68\sqr{2.1}3\sqr{1.5}3}

\def\CH{{\it CH }}

\newcommand{\myruleb}[1]{\vspace{0.5cm}\noindent\rule{#1}{0.05in}}
\newcommand{\myrulee}[1]{\newline\noindent\rule{#1}{0.05in\vspace{0.5cm}}}

\newenvironment{mybox}[3]
{\begin{minipage}{4.5in}
\setlength{\parindent}{0.5cm}
\myruleb{4.5in}
\label{box:#1}
\begin{center}\fbox{\textbf{#2}}\end{center}
{\begin{small}#3\end{small}}
\myrulee{4.5in}
\end{minipage}}

\newenvironment{myboxp}[4]
{\begin{minipage}{4.5in}
\setlength{\parindent}{0.5cm}
\myruleb{4.5in}
\label{box:#1}
\begin{center}\fbox{\textbf{#2}}\end{center}
\begin{center}\epsfig{figure=#3.ps,width=2in}\end{center}
{\begin{small}#4\end{small}}
\myrulee{4.5in}
\end{minipage}}

\newcommand{\myfig}[2]
{\begin{figure}[ht]
\begin{center}
\label{fig:#1}
\epsfig{figure=#1.ps,width=5in}
\caption{#2}
\end{center}
\end{figure}}

\newcommand{\myplotone}[2]
{\begin{figure}[ht]
\begin{center}
\label{fig:#1}
\plotone{#1.ps}
\caption{#2}
\end{center}
\end{figure}}

\newcommand{\myequn}[2]
{\begin{equation}
{#2}
\labequn{#1}
\end{equation}}

\newcommand{\myequna}[2]
{\begin{eqnarray}
{#2}
\labequn{#1}
\end{eqnarray}}

\newenvironment{headnoff}[1]
{\pnew\bHuge\bcenter{\bf\underline{#1}}\ecenter\eHuge}

\newcommand{\state}[1]{\noindent{#1}\vf}

\newcommand{\mnote}[1]{{{\small\it\marginpar{#1}}}}